\documentclass{aastex}     
\usepackage{spr-astr-addons}  
\usepackage{url}
\usepackage{color}
\usepackage{comment}
\usepackage{epstopdf}
\usepackage{gensymb}
\usepackage{xcolor}
\usepackage[caption = false]{subfig}
\usepackage{csvsimple, longtable, booktabs}
\usepackage{graphicx}

\newcommand{\bjdtdb}{\ensuremath{\rm {BJD_{TDB}}}}
\newcommand{\feh}{\ensuremath{\left[{\rm Fe}/{\rm H}\right]}}
\newcommand{\teff}{\ensuremath{T_{\rm eff}}}

\newcommand{\msun}{\ensuremath{\,M_\Sun}}
\newcommand{\rsun}{\ensuremath{\,R_\Sun}}
\newcommand{\lsun}{\ensuremath{\,L_\Sun}}
\newcommand{\mj}{\ensuremath{\,M_{\rm J}}}
\newcommand{\rj}{\ensuremath{\,R_{\rm J}}}
\newcommand{\fave}{\langle F \rangle}
\newcommand{\fluxcgs}{10$^9$ erg s$^{-1}$ cm$^{-2}$}

\usepackage[maxfloats=40] {morefloats} 

\begin{document}
\title{Optimization Applied to Selected Exoplanets}
\shorttitle{Optimization Applied to Selected Exoplanets}
\shortauthors{<Ng et al.>}

\author{Shi Yuan Ng\altaffilmark{1}}
\author{Zhou Jiadi\altaffilmark{2}}
\author{\c{C}a\u{g}lar P\"{u}sk\"{u}ll\"{u}\altaffilmark{3,4}}
\author{Timothy Banks\altaffilmark{5,6}}
\author{Edwin Budding\altaffilmark{7,8,9}}
\author{Michael D. Rhodes\altaffilmark{10}}

\altaffiltext{1}{DBS Risk Management Group, Model Validation, 12 Marina Boulevard, DBS Asia Central Level 14 @ MBFC Tower 3, Singapore 018982}
\altaffiltext{2}{Dept. Statistics \& Applied Probability, National University of Singapore, Blk S16, Level 7, 6 Science Drive 2, Singapore 117546}
\altaffiltext{3}{\c{C}anakkale Onsekiz Mart University, Faculty of Sciences and Arts, Physics Department, 17100 \c{C}anakkale, Turkey}
\altaffiltext{4}{Astrophysics Research Center and Ulup{\i}nar Observatory, \c{C}anakkale Onsekiz Mart University, 17100, \c{C}anakkale, Turkey}
\altaffiltext{5}{Nielsen, 200 W Jackson Blvd \#17, Chicago, IL 60606, USA. Email: tim.banks@nielsen.com, Tel: 1-847-284-4444}
\altaffiltext{6}{Physics \& Astronomy, Harper College, 1200 W Algonquin Rd, Palatine, IL 60067, USA}
\altaffiltext{7}{Dept. Physics \& Astronomy, UoC, Christchurch, New Zealand}
\altaffiltext{8}{SCPS, Victoria University of Wellington, Wellington, New Zealand}
\altaffiltext{9}{Carter Observatory, Wellington, New Zealand}
\altaffiltext{10}{Brigham Young University, Provo, Utah, USA}

\vspace{2mm}






	




\begin{abstract}

\noindent { Transit} and radial velocity models were applied to archival data in order to examine exoplanet properties, in particular for the recently discovered super-Earth GJ 357b.  There is however considerable variation in estimated model parameters across the literature, and especially their uncertainty estimates.  This applies even for relatively uncomplicated systems and basic parameters. Some published accuracy values thus appear highly over-optimistic.  We present our reanalyses with these variations in mind and specify parameters with appropriate confidence intervals for the exoplanets Kepler-1b, -2b, -8b, -12b, -13b, -14b, -15b, -40b \& -77b and 51 Peg.  More sophisticated models in WinFitter, EXOFAST, and DACE were applied, leading to mean planet densities for Kepler-12b, -14b, -15b, and -40b as: $0.11 \pm 0.01$, $4.04 \pm 0.58$, $0.43 \pm 0.05$, and $1.19^{+0.31}_{-0.36}$  g per cc respectively. We confirm a rocky mean density for the Earth-like GJ357b, although we urge caution about the modelling given the low S/N data.  We cannot confidently specify parameters for the other two proposed planets in this system. 
    
\end{abstract}


\keywords{Optimisation; Exoplanets; Light Curve Analysis; Radial Velocity Curve Analysis}


\section{Introduction}

In 1995 Mayor \& Queloz (1995) discovered an exoplanet orbiting the star 51 Peg.  Efforts over the following quarter of a century led to over four thousand\footnote{4,367 as of 17 March 2021.} `confirmed exoplanets' listed in the NASA Exoplanet Archive (NEA)\footnote{http://exoplanetarchive.ipac.caltech.edu/}, a clear indication of  the exponential interest in exo-planetary studies since this groundbreaking  paper. Out of these confirmed exoplanet discoveries, some three thousand were discovered using the transit method, making it currently the leading technique for exoplanet detection. In turn, the majority of the transit discoveries were based on data from the {\em Kepler} mission.  Borucki {\em et al.} (2003) set out the aims of the original {\em Kepler} mission within the context of exoplanet research, while a comprehensive early summary is that of Borucki et al.\ (2011). 

The Kepler Science Center manages the interface between the scientific mission and the {\em Kepler} data-using community. Importantly, data are freely and easily available from the NEA.  The {\em Kepler} data are some of the data sets available at this site.  Additionally the NEA provides tools for working with these data-sets, such as filtering and downloading selected quarters of data for a single specified {\em Kepler} target. The current paper primarily made use of \textit{Kepler} short cadence photometric data, extracted using the Python library \textit{lightkurve} (Cardos~{\em et al.}, 2018). Radial velocity data were also sourced from the NEA, with sources outlined below.

\begin{figure*}[htbp]
\centering
\includegraphics[width=6in]{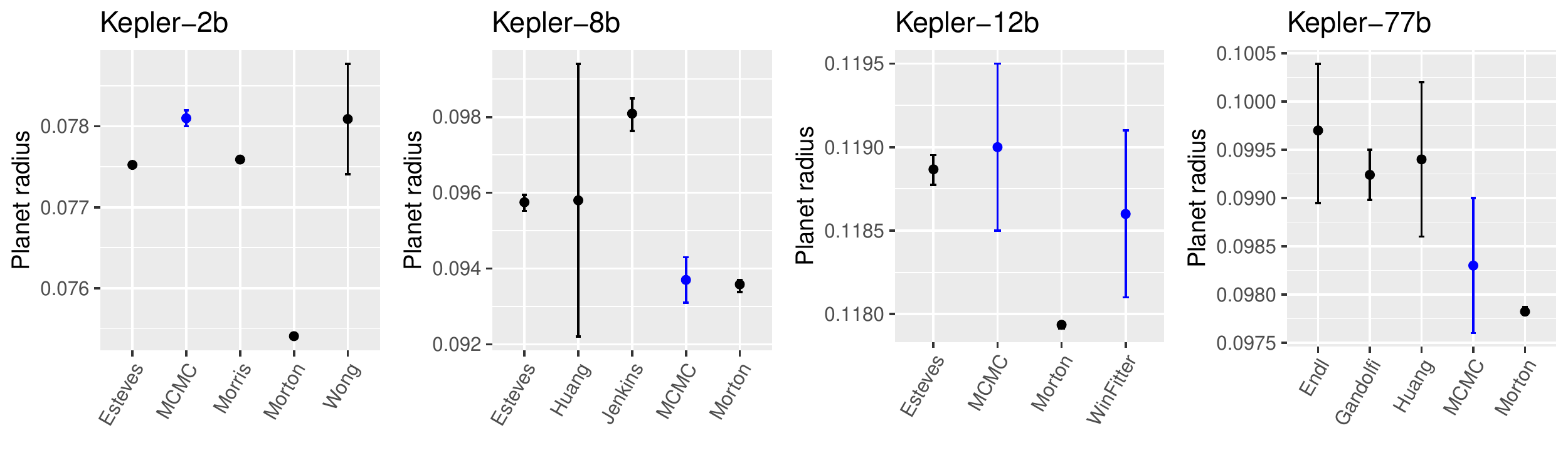}
\caption{
{\bf A Comparison of Planetary Radius Estimates (relative to their host star):} based on records in the NASA Exoplanet Archive (NEA) as an illustration of the uncertainties in the literature. Labels are first author surnames for the following papers: Endl~{\em et al.} (2015), Esteves~{\em et al.} (2015), Gandolfi~{\em et al.} (2011), Huang~{\em et al.} (2018), Jenkins~{\em et al.} (2010), Morris~{\em et al.} (2016), and Morton~{\em at al.} (2013). Further diagrams may be found as Figure 1 of Huang~{\em et al.} (2018). Blue points are from this paper: `MCMC' indicates the MCMC results (with one std. deviation error bars) with `WinFitter' being self-explanatory.
}
\label{fig:nea_comparison}
\end{figure*}

The paper's primary goal was to model to derive mean densities for several exoplanets, in particular the recently discovered super-earth system GJ357, and to have confidence in this analysis.  We therefore tested the models carefully using previously modelled systems, and report such results here both to show this testing but also since (as Figure~\ref{fig:nea_comparison} shows) there is a paucity of analyses for some of these systems and a general lack of consistency for many of them. Yet these are some of the `better' systems from the Kepler mission, with little or minor out-or-transit variations in intensity and deep, well-defined and well-sampled transits. We also made comparisons (see below) of the mean densities for previously modelled systems, to be confident in our analysis for GJ357.

{  This paper therefore will present comparisons for a number of systems, showing that uncertainties and difficulties (such as blended light, long integration times, sparsity of data by orbital phase, low signal to noise ratios, and host star photometric variability and difficulties in its `removal') in the collected data affect the confidence with which planetary parameters can be derived.} { On the basis of this we believe it is critical for the field} to establish the importance of realistic estimates for the uncertainties of exoplanet parameter specifications.   Fig~\ref{fig:nea_comparison} shows that literature values of even as basic a parameter as mean radius show considerable dispersion compared to published error estimates.  Parameters such as those associated with the limb-darkening of the host star are compromised all the more.   { Formal accuracies often seem to be over-estimated, to the point where different studies are producing estimates significantly different at high confidence levels.   We therefore recommend caution with the reliability of estimated parameters in the literature and} note that clarity of the parametrization is a necessary prelude to the meta-analysis that exoplanet research is moving towards, given the ongoing abundance of discoveries of exoplanets.  We believe that it is imperative to publish parameter estimates together with their uncertainties, as well as how these uncertainties were obtained, thereby indicating the reliability of these values.

The GJ 357 system is of particular relevance to the search for Earth-like planets.  However, the parametri\-zation depends on radial velocity data whose uncertainty is comparable to, if not greater than, the parameter values sought.  It is therefore important to have a clear view of the significance of findings in view of the presence of such measurement errors.

\begin{table*}[ht]
\centering
\caption{
{\bf Comparison of point estimates from fitting of radial velocity data for 51 Peg-b, Kepler-12b, -14b, -15b, and -40b.} 
GA stands for results using the genetic optimisation method in this paper, and LM for results using the  Levenberg--Marquardt method.
$e$ is eccentricity, $RV_C$ is systemic velocity, and $q$ the mass ratio. 
The parameters are taken from the same papers as the data. The 51 Peg data are from Butler~{\em et al.}
(2006), but the parameters given are from Bedell~{\em et al.} (2019). { Papers are referred to by lead author name: Bedell~{\em et al.} (2019), Buchave~{\em et al.} (2011), Butler~{\em et al.} (2006), Endl~{\em et al} (2011), Fortney~{\em et al.} (2011), Rosenthal et al. (2021), and Santerne~{\em et al.} (2011).
\label{table:rv_fits_single}
The derived estimates from these studies of the Kepler systems have been adopted as foundational by other researchers such as Southworth (2012) and Esteves et al. (2015), being the only papers modelling these radial velocity data.  The NEA lists 5 papers with radial velocity solutions for 51 Peg, those selected had included eccentricity as a free parameter and so were directly comparable to the fit by this study. }
}
\begin{tabular}{llllll}
\hline
System & Method        & $e$    & $RV_C$ (km/s)  & $ q $ $ (10^{-4})$  & Semi-amplitude (m/s) \\
\hline
Kepler 12b   & LM             & 0.11   & 0.000          & 3.94                 & 54.1 \\
Kepler 12b   & GA             & 0.36   & 0.376          & 3.84                 & 56.1 \\
Kepler 12b   & Fortney        & $0.00^{+0.01}_{-0.01}$  & $0.0792^{+0.0071}_{-0.0070} $         & $3.53^{+0.52}_{-0.46}$                 & $48.2^{+4.4}_{-4.3}$ \\
\hline
Kepler 14b   & LM             & 0.03   & 2.490          & 32.4                 & 404.4 \\
Kepler 14b   & GA             & 0.09   & 1.671          & 32.6                 & 409.0 \\
Kepler 14b   & Buchave        & $0.035^{+0.020}_{-0.020}$   & $6.53^{+0.30}_{-0.30}$           & $53.00^{+3.83}_{-3.55}$                 & $682.9^{+26.7}_{-24.6}$ \\
\hline
Kepler 15b   & LM             & 0.16   & 24.02          & 6.35                 & 83.5 \\
Kepler 15b   & GA             & 0.26   & 18.95          & 6.38                 & 85.7 \\
Kepler 15b   & Endl           & ---    & $20.0^{+1.0}_{-1.0}$           & $6.19^{+1.12}_{-1.06}$                 & $78.7^{+8.5}_{-9.5}$ \\
\hline
Kepler 40b   & LM             & 0.00   & 6.591          & 17.0                 & 220 \\
Kepler 40b   & GA             & 0.14   & 6.582          & 7.0                  & 100 \\
Kepler 40b   & Santerne       & 0.00 (fixed)   & $6.565^{+0.020}_{-0.020}$          & $14.20^{+3.29}_{-3.03}$                 & $179^{+27}_{-27}$ \\
\hline 
51 Peg-b & LM          & 0.021   & 0.000         & 4.30                 & 56.6 \\
51 Peg-b & GA          & 0.021   & 0.295         & 4.04                 & 53.1 \\
51 Peg-b & Bedell      & $0.03^{+0.02}_{-0.02}$    & ---           & ---                 & $55.57^{+2.28}_{-2.04}$ \\
51 Peg-b & Butler & $0.013 \pm 0.12$ & --- & --- & $55.94 \pm 0.69$ \\
51 Peg-b & Rosenthal   & $0.0042^{+0.0046}_{-0.0030}$ & ---  & --- & $55.73^{+0.32}_{-0.30}$ \\
\hline
\end{tabular}
\end{table*}


\begin{table*}[!htb] 
\caption{\label{table:single_transit} {\bf Transit Model Fits} using the LM and GA optimisation methods, compared with selected published and {\sc WinFitter} (WF) results. NEA is a representative paper from the NEA tables. Inclination $i$ is in degrees, $k$ is the planet radius ($R_P$) divided by the stellar radius ($R_S$), and  $a$ is the semi-major axis of the exoplanet's orbit. Values of limb darkening denoted with * were linearly interpolated using tables from Sing (2010).
}
{\footnotesize
\centering
\begin{tabular}{llllllll}
\hline
Planet     & Source & $k$                          & ${R_{S} / }{a}$              & $i$                     & $u$                        & $u_1$                        & $u_2$                        \\ \hline
Kepler-1b  & LM     & 0.1272                       & 0.1278                       & 83.77                   & 0.6709                     & ---                          & ---                          \\
Kepler-1b  & LM     & 0.1254                       & 0.1294                       & 83.67                   & ---                        & 0.3913                       & 0.3308                       \\
Kepler-1b  & GA     & 0.1245                       & 0.1275                       & 83.72                   & 0.4852                     & ---                          & ---                          \\
Kepler-1b  & GA     & 0.1244                       & 0.1275                       & 83.72                   & ---                        & 0.3516                       & 0.1892                       \\
Kepler-1b  & NEA    & $0.1254^{+0.0005}_{-0.0004}$ & $0.1265^{+0.0003}_{-0.0004}$ & $83.87^{+0.02}_{-0.02}$ & 0.5600                    & $0.3300^{+0.0800}_{-0.0600}$ & $0.2850^{+0.0620}_{-0.0870}$ \\
Kepler-1b  & WF     & $0.1268 \pm 0.0247$ & $0.1251 \pm 0.0117$  &    $84.00 \pm 0.81$ & $0.6485 \pm 0.0056 $  & ---                          & ---                          \\ \hline
Kepler-2b  & LM     & 0.0781                       & 0.2476                       & 82.22                   & 0.4635                     & ---                          & ---                          \\
Kepler-2b  & LM     & 0.0776                       & 0.2463                       & 82.45                   & ---                        & 0.3390                       & 0.2003                       \\
Kepler-2b  & GA     & 0.0777                       & 0.2412                       & 82.98                   & 0.4696                     & ---                          & ---                          \\
Kepler-2b  & GA     & 0.0774                       & 0.2417                       & 82.98                   & ---                        & 0.3689                       & 0.1642                       \\
Kepler-2b  & NEA    & $0.0775 \pm 0.0001$          & $0.2407^{+0.0001}_{-0.0002}$ & $83.14 \pm 0.02$        & 0.5100                     & $0.3497^{+0.0026}_{-0.0035}$ & $0.1741^{+0.0057}_{-0.0044}$ \\
Kepler-2b  & WF     & $0.0780 \pm 0.0001 $         & $0.2347 \pm 0.0001$          & $83.94 \pm 0.02$        & $0.4871 \pm 0.0054$                        & ---                          & ---                          \\ \hline
Kepler-8b  & LM     & 0.0943                       & 0.1531                       & 83.42                   & 0.5488                     & ---                          & ---                          \\
Kepler-8b  & LM     & 0.0939                       & 0.1531                       & 83.43                   & ---                        & 0.4752                       & 0.1040                       \\
Kepler-8b  & GA     & 0.0946                       & 0.1557                       & 83.20                   & 0.6189                     & ---                          & ---                          \\
Kepler-8b  & GA     & 0.0917                       & 0.1432                       & 84.23                   & ---                        & 0.2971                       & 0.3394                       \\
Kepler-8b  & NEA    & $0.0958 \pm 0.0002$          & $0.1459 \pm 0.0004$          & $83.98^{+0.04}_{-0.03}$ & 0.5200                    & $0.3710^{+0.0390}_{-0.0560}$ & $0.1610^{+0.0710}_{-0.0510}$ \\
Kepler-8b  & WF     & $0.0978 \pm 0.0001 $         & $0.1646 \pm 0.0001$          & $ 82.53 \pm 0.01$       & 0.5477*                        & ---                          & ---                          \\ \hline
Kepler-12b & LM     & 0.1192                       & 0.1303                       & 87.35                   & 0.4912                     & ---                          & ---                          \\
Kepler-12b & GA     & 0.1171                       & 0.1241                       & 88.92                   & 0.5173                     & ---                          & ---                          \\
Kepler-12b & GA     & 0.1179                       & 0.1247                       & 88.80                   & 0.5700*                        & 0.4699                       & 0.0535                       \\
Kepler-12b & NEA    & $0.1189 \pm 0.0001$          & $0.1247 \pm 0.0002$          & $88.80^{+0.09}_{-0.07}$ & 0.6050                    & $0.4205^{+0.0060}_{-0.0058}$ & $0.1370 \pm 0.0130$          \\
Kepler-12b & WF     & $0.1186 \pm 0.0005 $         & $0.1240 \pm 0.0014 $         & $ 88.91 \pm 0.57 $      & $ 0.4741 \pm 0.0154 $      & ---                          & ---                          \\ \hline
Kepler-13b & LM     & 0.0647                       & 0.2541                       & 81.94                   & 0.5812                     & ---                          & ---                          \\
Kepler-13b & GA     & 0.0700                       & 0.2315                       & 82.69                   & 0.4302                     & ---                          & ---                          \\
Kepler-13b & NEA    & $0.0874 \pm 0.0001$          & $0.2222 \pm 0.0002$          & $86.77 \pm 0.05$        & 0.5330                    & $0.3183^{+0.0020}_{-0.0021}$ & $0.2024 \pm 0.0042$          \\ 
Kepler-13b & WF & $0.0857 \pm 0.0001$ & $0.2250 \pm 0.0001$ & $85.85 \pm 0.17$ & 0.5602* & --- &  --- \\ \hline
Kepler-14b & LM     & 0.0590                       & 0.1403                       & 84.81                   & 0.4299                     & ---                          & ---                          \\
Kepler-14b & GA     & 0.0573                       & 0.1125                       & 91.60                   & 0.4632                     & ---                          & ---                          \\
Kepler-14b & GA     & 0.0579                       & 0.1293                       & 85.99                   & ---                        & 0.2880                       & 0.2528                       \\
Kepler-14b & NEA    & $0.0569 \pm 0.0013$          & $0.1218^{+0.0014}_{-0.0080}$ & $90.00^{+0.00}_{-2.80}$ & 0.5600                    & 0.3070                      & 0.3130                      \\
Kepler-14b & WF     & $ 0.0455 \pm 0.0004 $        & $0.1334 \pm 0.0023 $         & $85.54 \pm 0.24 $       & $ 0.5095 \pm 0.0519$       & ---                          & ---                          \\ 
Kepler-14b & WFG  & $ 0.0422 \pm 0.0006 $        & $0.1139 \pm 0.0059 $         & $87.92 \pm 1.10 $         & $ 0.4869 \pm 0.4406 $ & --- & --- \\ \hline
Kepler-15b & LM     & 0.0992                       & 0.1053                       & 85.67                   & 0.6246                     & ---                          & ---                          \\
Kepler-15b & GA     & 0.0936                       & 0.0880                       & 87.28                   & 0.6630                     & ---                          & ---                          \\
Kepler-15b & GA     & 0.0910                       & 0.0851                       & 92.24                   & ---                        & 0.5194                       & 0.3428                       \\
Kepler-15b & NEA    & $0.0996^{+0.0006}_{-0.0005}$ & $0.0781^{+0.0104}_{-0.0067}$ & $87.44^{+0.18}_{-0.20}$ & 0.6270                    & 0.4310                      & 0.2430                      \\
Kepler-15b & WF     & $ 0.1008 \pm 0.0015 $        & $ 0.0935 \pm 0.003  $        & $ 86.80 \pm 0.33 $      & $ 0.6525 \pm 0.0586$       & ---                          & ---                          \\ \hline
Kepler-77b & LM     & 0.0983                       & 0.1044                       & 87.63                   & 0.5793                     & ---                          & ---                          \\
Kepler-77b & LM     & 0.0960                       & 0.1023                       & 88.18                   & ---                        & 0.3623                       & 0.4841                       \\
Kepler-77b & GA     & 0.0970                       & 0.1025                       & 92.00                   & 0.6210                     & ---                          & ---                          \\
Kepler-77b & GA     & 0.0964                       & 0.1010                       & 88.31                   & ---                        & 0.5040                       & 0.2088                       \\
Kepler-77b & NEA    & $0.0992 \pm 0.0003$          & $0.1024 \pm 0.0006$          & $88.00 \pm 0.11$        & 0.6780                    & $0.5050 \pm 0.0150$          & $0.1390 \pm 0.0320$          \\
Kepler-77b & WF     & $0.0994 \pm 0.0003$          & $0.1019 \pm 0.0003$          & $88.09 \pm 0.05$        & 0.6780*                        & ---                          & ---                          \\ 
Kepler-77b & WFG   & $0.0968 \pm 0.0005$ & $0.1015 \pm 0.0018$ & $86.14 \pm 0.34 $ & 0.6780* & --- & --- \\ \hline
\end{tabular}
}
\end{table*}

\begin{table*}[!htb]
\caption{\bf \label{table1} Primary Input Data for Transit Curve Fitting by {\sc WinFitter}.}
\centering
\scalebox{0.9}
{
	\begin{tabular}{cccccclll}
	\hline 
	System & 
	$M_\star (M_{\odot})$ & $R_\star (R_{\odot})$ & $T_\star (K \deg )$ & $T'_{eq} (K \deg)$ & $u$ & Epoch (BJD) & $P (days)$ & Reference \\
	\hline 
	\noalign{\smallskip}
	\mbox{Kepler-12} & 1.166 & 1.483 & 5947 & 1480 & 0.57 & 2455004.00915 & 4.4379629 & Esteves {\em et al.}, 2015 \\ 
	\mbox{Kepler-14} & 1.512 & 2.048 & 6395 & 1605 & 0.53 & 2454971.08737 & 6.7901230 & Buchhave {\em et al.}, 2011 \\ 
	\mbox{Kepler-15}& 1.018 & 0.992 & 5515 & 1251 & 0.64 & 2454969.328651 & 4.942782 & Endl {\em et al.}, 2011 \\ 
	 \hline 
	\end{tabular}
}
\vspace{1ex}
\end{table*} 


\section{Method}

The overall flow of the project was:
\begin{itemize}
\item{Build a simple-planetary transit light curve and radial velocity models using the Python and R programming languages.}

\item{Fit the models using simple optimizers (such as the Levenberg-Marquadt algorithm).

\item{Apply more sophisticated modelers such as {\sc WinFitter}  (described further below), comparing results from those from the previous steps and with the literature,} to see if similar results were obtained.  Should there be good agreement, we would move on to the next step.} 

\item{Employ Markov Chain Monte Carlo (MCMC) procedures to obtain estimates and uncertainties of the parameters for the systems.}

\item{Derive density estimates for those systems with both radial velocity and transit fits.  Use the more sophisticated EXOFAST modeler (described below) to fit simultaneously the radial velocity and transit data sets, and compare results.}

\end{itemize}
Our goals were to build and test models, first using synthetic data and then for systems with published analyses, so that we could confirm that both the models and the optimisation techniques were resulting in reasonable estimates.  This would lend confidence to the later analyses, where we applied the more time consuming MCMC methodologies to estimate the planetary densities for a number of systems. Publicly available data were used in this study, sourced (as noted below) from the literature, the MAST data archive at the Space Telescope Science Institute (STScI), or the NEA. { A wide variety of modelling programs are used in the exoplanet literature, making a comparison of all such tools a substantial task.  Instead we selected several of the more popular techniques for direct comparison in this paper, together with comparisons of the literature at the system level.}

\subsection{Build Initial Models}

A light curve model for photometric data and a radial velocity model for spectroscopic data were built (in Python) from first principles using Mandel~\& Agol's (2002) and Budding~\& Demircan's (2007) formulations for transit modelling, and Haswell (2010) \& Hatzes (2016) for the radial velocity models. Orbital eccentricity was accounted for in the radial velocity model and limb darkening laws (linear and quadratic) were used in the photometric model.  The `small planet' approximation was used for the transit model, in that the limb darkening value/s corresponding to the centre of the planetary disk projected onto the stellar disk were uniformly applied across the stellar area obscured by the planet.  Heller (2019) reported that the effect of this approximation is an order of magnitude less that uncertainties arising from the total noise budget in light curves, translating into typical errors in the derived planetary radius ($R_{P}$) of $\sim 10^{-4}$ for $R_{P} =0.1$ in highly accurate space based observations of bright stars (Gilliland {\em et al.}, 2011) such as used in this study. Heller notes that the approximation produces errors orders of magnitude smaller than the error coming from uncertain limb darkening coefficients.  Croll {\em et al.} (2007) note that the small planet approximation is valid for $\frac{R_P}{R_S} < 0.1$, where $R_S$ is the stellar radius.  The systems we looked are generally below this limit, however two of the transit test cases exceeded it and hence one explanation why we also used more sophisticated models later in the paper.

\begin{table*}[!htb] 
\caption{\label{table:mcmc_transits } {\bf MCMC Transit Model Fits.} Errors are one standard deviation. $i$ is in degrees, $k$ is the ratio of the planetary to stellar radius, $R_S$ is the stellar radius, and $u$ is the linear limb darkening
coefficient. Data were not yet corrected via Gaussian Processes for Kepler-14 and -77.}
\centering
\begin{tabular}{lrllrl}
\hline
Planet      & $k$  $ (10^{-3})$                & ${R_{S} / }{a}$   $ (10^{-3})$    & $u$                 & $cos(i)$  $ (10^{-3})$            &  $i$ (degrees)  \\
\hline
Kepler-1b   & $ 126.8 \pm 0.3$ & $ 127.7 \pm 0.3$ & $0.648 \pm 0.025$ & $108.4  \pm 0.4$  &  $83.77  \pm 0.02$\\
Kepler-2b   & $  78.1 \pm 0.1$ & $ 247.3 \pm 1.2$ & $0.468 \pm 0.005$ & $134.5  \pm 2.3$  &  $82.27  \pm 0.13$\\
Kepler-8b   & $  93.7 \pm 0.6$ & $ 148.8 \pm 3.3$ & $0.513 \pm 0.043$ & $108.8  \pm 0.5$  &  $83.75  \pm 0.03$\\
Kepler-12b  & $ 119.0 \pm 0.5$ & $ 129.1 \pm 1.5$ & $0.495 \pm 0.005$ &  $42.1  \pm 0.5$  &  $87.59  \pm 0.03$\\
Kepler-13b  & $  65.6 \pm 0.1$ & $ 310.8 \pm 6.1$ & $0.461 \pm 0.007$ & $227.9  \pm 10.2$ &  $76.83  \pm 0.60$\\
Kepler-14b  & $  57.9 \pm 0.6$ & $ 133.2 \pm 6.3$ & $0.465 \pm 0.033$ &  $77.2  \pm 11.4$ &  $85.57  \pm 0.66$\\
Kepler-15b  & $  99.6 \pm 0.4$ & $ 106.7 \pm 1.3$ & $0.610 \pm 0.024$ &  $77.9  \pm 1.9$  &  $85.53  \pm 0.11$\\
Kepler-77b  & $  98.3 \pm 0.7$ & $ 107.0 \pm 0.2$ & $0.565 \pm 0.021$ &  $48.5  \pm 0.5$  &  $87.22  \pm 0.03$\\
\hline
\label{table:mcmc_transits}
\end{tabular}
\end{table*}


\begin{table*}[!htb]
\caption{{\bf MCMC Transit Model Fits for data following subtraction of out of transit flux variations, using Gaussian Process models.} Errors are one standard deviation. $i$ is in degrees, $k$ is the ratio of the planetary to stellar radius, $R_S$ is the stellar radius, and $u$ is the linear limb darkening.}
\centering
\begin{tabular}{lrllrl}
\hline
Planet      & $k$  $ (10^{-3})$                & ${R_{S} / }{a}$   $ (10^{-3})$    & $u$                 & $cos(i)$  $ (10^{-3})$            &  $i$ (degrees)  \\
\hline
Kepler-14b & $46.8 \pm 0.3$ & $148.8 \pm 5.0 $ & $0.477 \pm 0.027$ & $ 102.5 \pm 7.4 $ & $ 84.12 \pm 0.43$ \\
Kepler-77b & $98.6 \pm 0.5$ & $104.5 \pm 1.6 $ & $0.579 \pm 0.140$  & $ 41.4 \pm 0.4$   & $ 87.37 \pm 0.25$ \\
\hline
\label{table:gp_mcmc_transits}
\end{tabular}
\end{table*}


\begin{table*}[!htb]
\caption{\label{table:mcmc_transits_rhat } {\bf MCMC Transit Model Parameter $\hat{R}$ values}}
\centering
\begin{tabular}{p{1cm}p{1.5cm}p{1.5cm}p{1.5cm}p{1.5cm}p{1.5cm}p{1.5cm}p{1.5cm}p{1.5cm}}
\hline
 & Kepler-1b & Kepler-2b & Kepler-8b & Kepler-12b & Kepler-13b & Kepler-14b & Kepler-15b & Kepler-77b \\
\hline
$k$ & 1.0001 & 1.0023 & 0.9997 & 1.0050 & 1.0008 & 1.0001 & 0.9999 & 0.9999 \\
$R_S/a$ & 1.0021 & 1.0007 & 1.0001 & 1.0049 & 1.0005 & 1.0002 & 0.9997 & 1.0003 \\
$u$ & 1.0013 & 1.0000 & 0.9998 & 1.0021 & 0.9996 & 0.9996 & 1.0013 & 1.0002 \\
$cos(i)$ & 1.0025 & 1.0009 & 1.0002 & 1.0060 & 1.0004 & 1.0004 & 0.9997 & 1.0004 \\
\hline
\label{table:mcmc_transits_rhat}
\end{tabular}
\end{table*}

 
 \begin{table*}[!htb] 
\caption{\label{table:mcmc_radial } {\bf MCMC Radial Velocity Model Fits.} Errors are one standard deviation. $RV_C$ is in km per second.}
\centering
\begin{tabular}{lllr}
 \hline
 Planet     & $e$                 & $q$  $(10^{-5})$                     & $RV_C$ \\
 \hline
 Kepler-12b & $0.1594 \pm 0.0059$ & $36.224 \pm 0.085$ & $4.4167 \pm 0.1849$ \\
 Kepler-14b & $0.0373 \pm 0.0008$ & $324.183 \pm 0.124$  & $0.8351 \pm 0.1287$ \\
 Kepler-15b & $0.1852 \pm 0.0031$ & $64.141 \pm 0.112$  & $16.1181 \pm 0.1338$ \\
 Kepler-40b & $0.3193 \pm 0.0077$ & $0.139 \pm 0.001$  & $6.5500 \pm 0.0005$ \\
 51 Peg-b & $0.0188 \pm 0.0004$ & $42.349 \pm 0.009$  & $-2.4585 \pm 0.0099$ \\
\hline
\end{tabular}
\label{table:mcmc_radial}
\end{table*}


\subsection{Model Tests}

Checks were made that the models give parameter estimates in line with the literature.  Two different optimisers (the Genetic and the Levenberg-Marquardt (LM) algorithms) were applied to known exoplanetary systems:
\begin{itemize}
    \item{Radial velocity data for 51 Peg-b, Kepler-12, -14, -15, and -40. The 51 Peg data are from Butler~{\em et al.} (2006).  Kepler-12 data are from Fortney~{\em at al.}. Kepler-15 data are from Endl~{\em et al} (2011), and Kepler-14 from Buchave~{\em et al.} (2011).  The Kepler-14b fits below correspond to Buchave~{\em et al.'s} `uncorrected' fit in that dilution of the host star's light by the nearly equal magnitude stellar companion ($\approx$ 0.5 mag fainter) was not taken into account. Kepler-40 data were taken from Santerne~{\em et al.} (2011), }
    \item{Photometric data from Kepler, sourced from the NEA. Quarter 1 data was used for Kepler-1b. For Kepler-2b, -12b and -13b, quarter 2 data were extracted. Quarter 3 data were utilized for Kepler-14b and -15b, whereas quarter 5 data was used for Kepler-8b and -77b. We selected quarters that, by visual inspection, minimised out of transit variations in flux. We note that Kepler-1b is also known as TrES-2b, being discovered by O'Donovan~{\em et al.} (2006). Most of the systems chosen were selected because they had no out-of-transit effects, for instance, Kepler-1 has a uniform flux outside of the transit region.  We did not analyse the entire 3 years of Kepler data due to its volume and lack of variation, besides these were essentially test cases to validate that the software was producing results in line with the literature. Two of the systems (Kepler-14 and 77) exhibited waves running through the out-of-transit data, which we subsequently modelled using Gaussian Processes (GP) to `clean' and remove these disturbances before the transits were modelled. A GP is a collection of finite number random variables, which have a joint Gaussian distribution (Rasmussen \& Williams, 2005). This means that a GP can be completely determined by its mean function and covariance function. In order to remove the stellar variability in the transit, a GP model is first fitted to the out-of-transit data and then used for prediction. To facilitate a smoother light curve fitting, we remove these predictions from the original data, which includes the transit data. We made use of the {\em Juliet} package (Espinoza~{\em et al.}, 2019).}
\end{itemize}

The LM algorithm can be seen as a combination of the steepest gradient algorithm and the Newton algorithm (Li {\em et al.}, 2016).  In this project, the LM algorithm is implemented using the \textit{nls.lm} function from the \textit{minpack.lm} package in R (Timur {\em et al.}, 2016). LM gives standard error estimates, although these formal errors were clearly too small by several factors of ten. We therefore do not report these errors as we place little confidence in them, instead preferring to later use Monte Carlo methods to make such estimates. 

The R package {\em genalg}\footnote{https://github.com/egonw/genalg} was used to implement the genetic optimisations.

The other code we have applied for light curve analysis is graphical user interfaced close binary system analysis program, {\sc Winfitter}. This program is described in Rhodes~\& Budding (2014). It uses a modified Levenberg-Marquardt optimisation technique to find the model light curve that corresponds to the least value of $\chi^2 $. The fitting function is based on the Radau formulations of Kopal (1959). These formulations allow analysis of the tidal and rotational distortions (ellipticity), together with the radiative interactions (reflection), for massive and relatively close gravitating bodies (a detailed background given on Budding~{\em et al.}, 2016). WinFitter has several interesting points: 
\begin{itemize}
    \item{The relatively simple and compact algebraic form of the fitting function, which allows large regions of the  $\chi^2$  parameter space to be explored at low computing cost. }
    \item{The $\chi^2$ Hessian (see, for example, Bevington, 1969) can be simply evaluated in the vicinity of the derived minimum. Inspecting this matrix, and in particular its eigenvalues and eigenvectors, gives valuable insights into parameter determinacy and interdependence.}
    \item{The Hessian can be inverted to yield an error matrix. This must be positive definite if a determinate, `unique' optimal solution is to be evaluated. WinFitter considers strict application of this provision essential to avoiding over-fitting the data.}
\end{itemize}
The program uses three different optimisation techniques:
\begin{enumerate}
    \item{parabolic interpolation for single parameters, in a step by step mode;}
    \item{Parabolic in conjugate directions; and}
    \item{ Vector (in all parameters) mode.}
\end{enumerate}
The program switches between these modes depending on the convergence rate and user defined limits.  Further details on how WinFitter addresses the information content of data and estimation of uncertainties can be found in Banks \& Budding (1990).

The first step in optimisation used as free parameters, $ k, R_s/a, i$ and $u$, and then was followed by a second step fixing these initial parameters to their derived values and optimising for for $e$ and $M_0$.  Inclination is denoted as $i$, $k$ is the planet radius ($R_P$) divided by the stellar radius ($R_S$), $u$ is the linear limb darkening value, and  $a$ is the semi-major axis of the exoplanets orbit.   The parameter errors were derived from the Hessian inverse matrix calculated at the adopted optimum. The formal error estimates are described in detail by Budding~{\em et al.} (2016b).


\subsubsection{Radial Velocity Tests}

Table~\ref{table:rv_fits_single} lists the results for the radial velocity fits, showing the LM fits to be in general agreement with the papers modeling the { \bf same} data,\footnote{ Bar Butler {\em et al.} (2006) and Rosenthal {\em et al.} (2021) which were included to comfirm that other researchers had found indication of eccentricity as well.} and  confirming the usefulness of this paper's simple model.  The genetic algorithm tended to settle on more divergent solutions, longer run times might have lead to improvement through `wider' searches. Here, we note that the value for $q$ and its associated errors for 51 Peg could not be obtained from Bedell~{\em et al.} (2019) due to non-specification of $M_*$. Additionally, an outlier $(BJD=2455019.11155,RV=36.7,\sigma_{RV}=19.4)$ was removed from data for Kepler-12b before usage for analysis. This was done to minimise disruption of the sine wave structure for the radial velocity model so that a better overall fit could be obtained. 


\subsubsection{Transit Tests}

\begin{figure*}[htbp]
\centering
\subfloat[]{\includegraphics[width=3in]{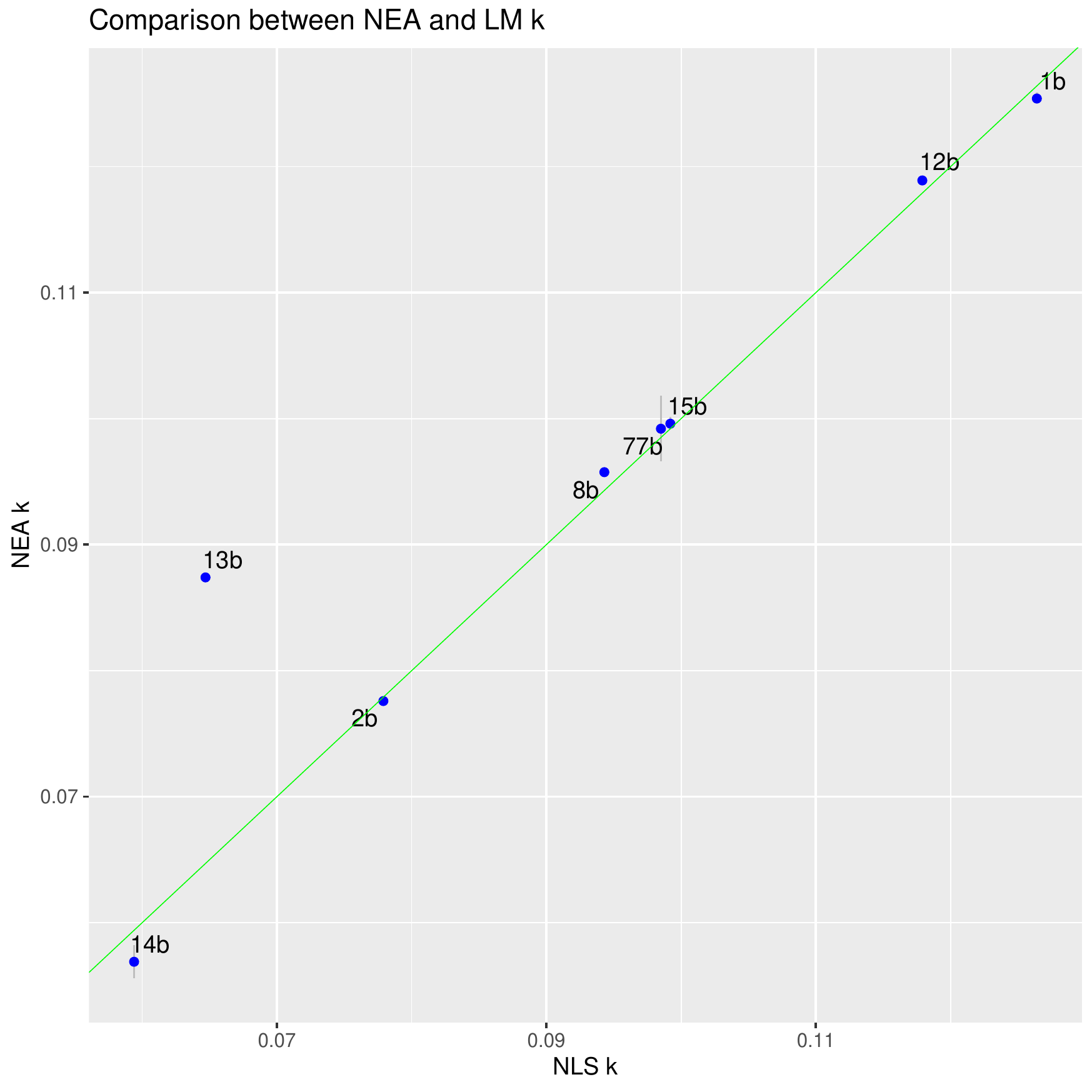}}
\subfloat[]{\includegraphics[width=3in]{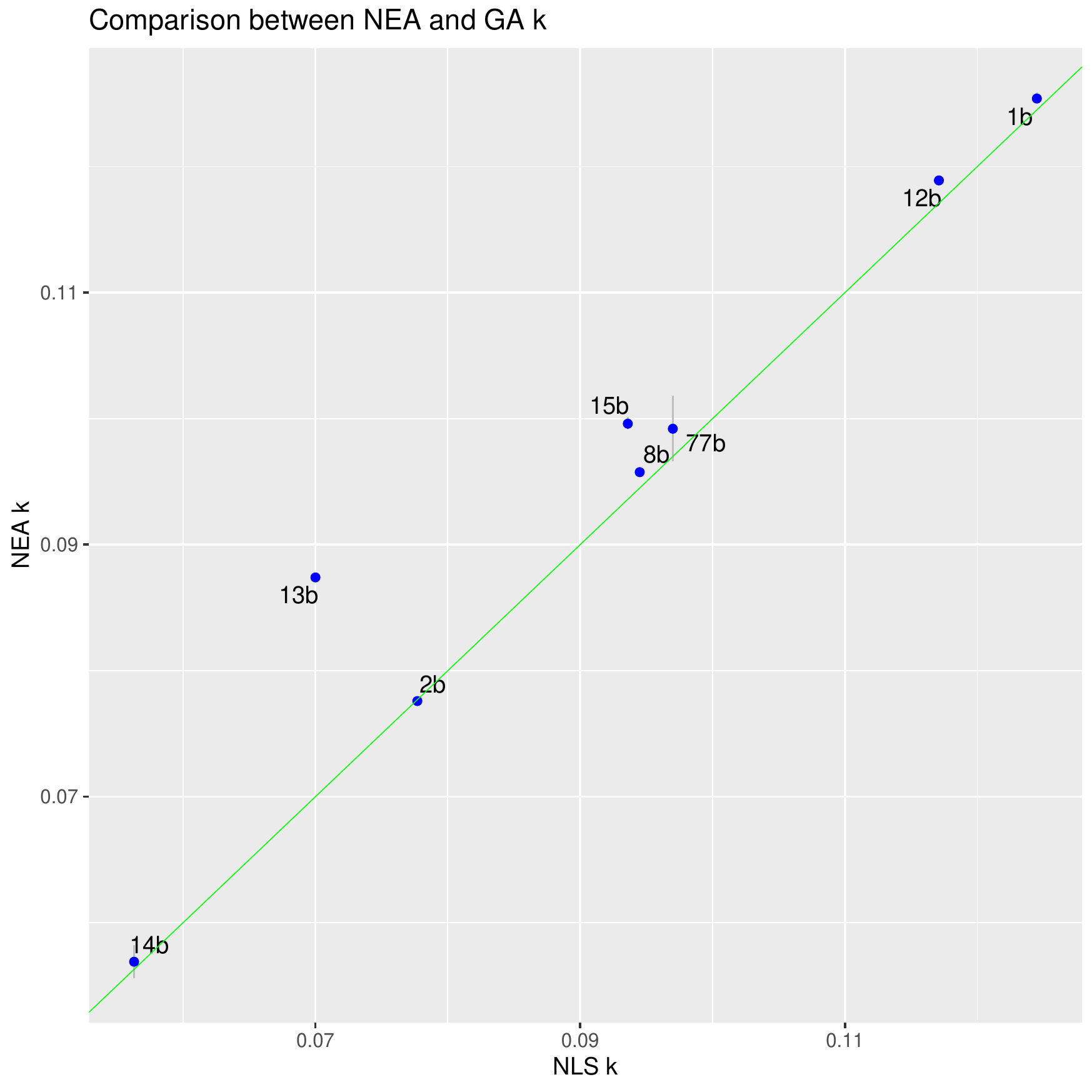}} \\
\subfloat[]{\includegraphics[width=3in]{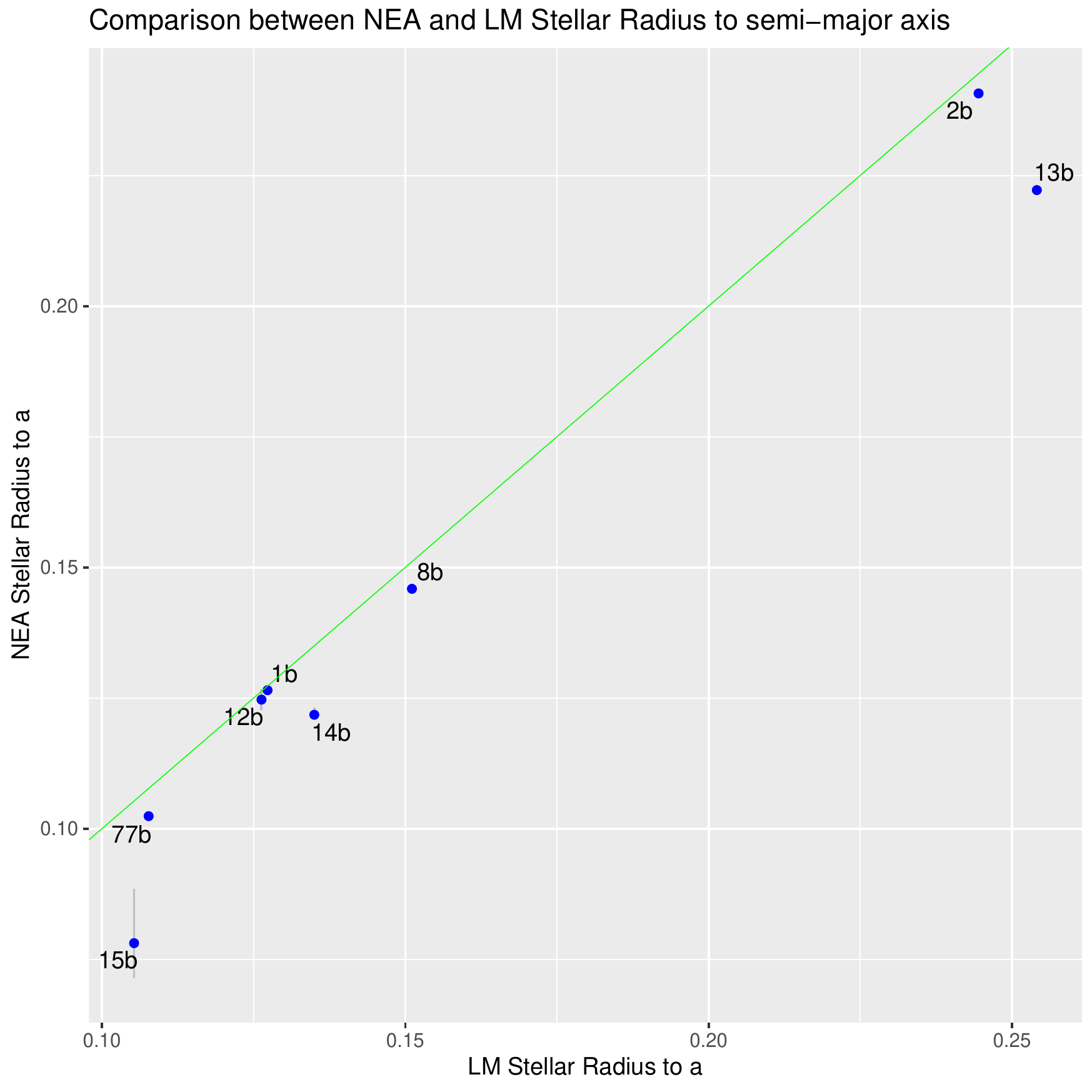}}
\subfloat[]{\includegraphics[width=3in]{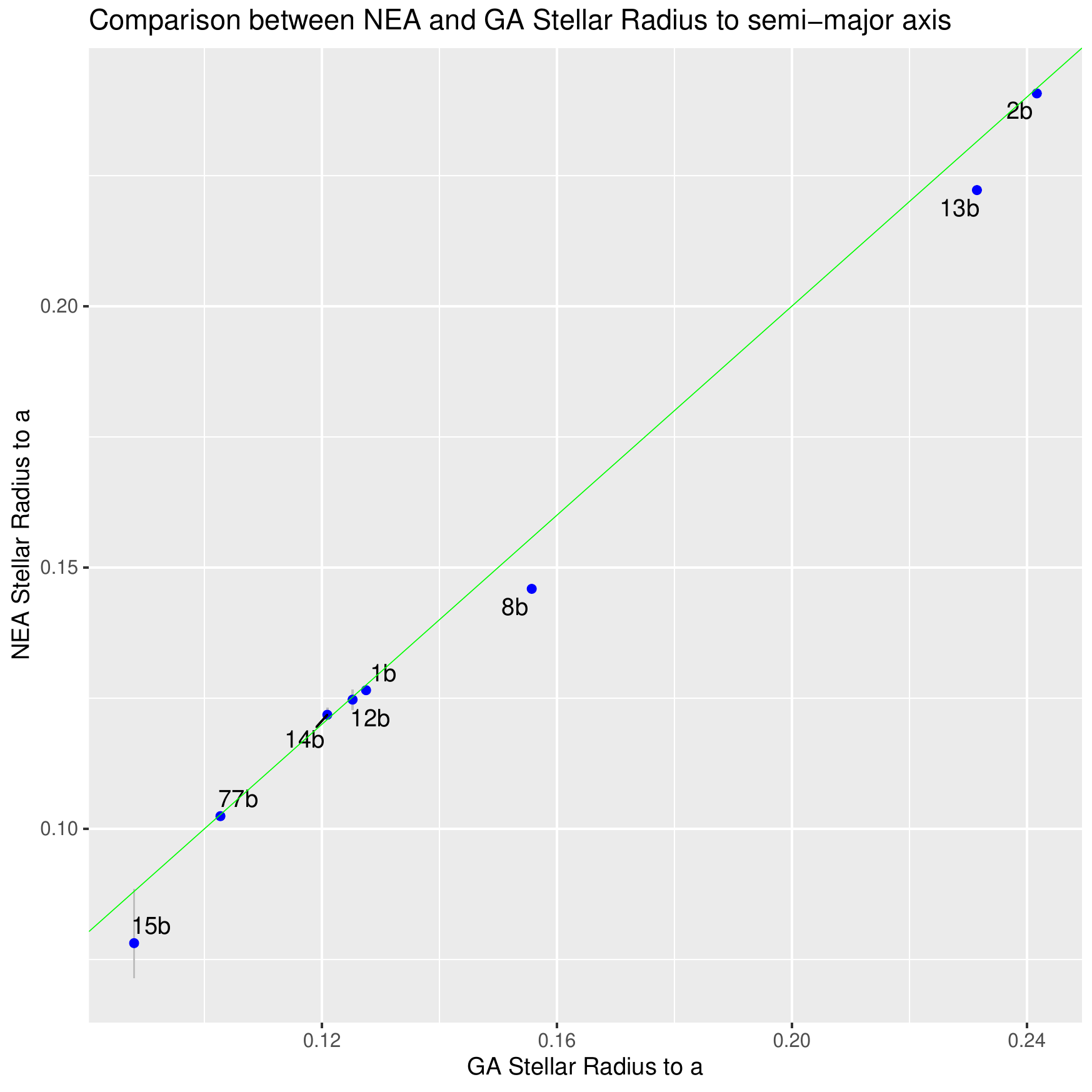}} 
\caption{{{\bf Comparison of Genetic Algorithm, LM, and NEA results:} the left hand column compares LM
results against the NEA-based figures for $k$ and$\frac{r_s}{a}$ and the right column compares the same variables for
the GA method against NEA-based results. The green lines indicate where both methods plotted would be in agreement. These sample charts show the physical ranges being tested across and the general agreement of the point estimates with the NEA-recommended values.
}}
\label{fig:nea_comparison2}
\end{figure*}

Table~\ref{table:single_transit} gives results for the photometric modeling, showing again general agreement across the  different methods or sources {(see also Fig~\ref{fig:nea_comparison2} which displays a comparison with the NEA `preferred solutions' across the systems).} For Kepler-13b, the model used to obtain these estimates allowed eccentricity to be a free parameter. For all other systems, eccentricity was fixed at 0. This is as Kepler-13b is a system with complex effects which are further explored in Budding~{\em et al.} (2018). Such treatment would give more reliable estimates since lesser assumptions are made about the system, although we acknowledge that it is a limited representation of the effects present. In most fits only linear limb darkening could be constrained ($u$), only in one fit for Kepler-13b were the quadratic coefficients ($u_1$ and $u_2$) obtainable. {\sc Winfitter} refers to a model fit using that program, and used Claret~\& Bloemen (2011) to determine $u$ (based on literature information as given in Table~\ref{table1}). Formal errors are not given in this table, but are discussed in the MCMC fitting results.  The goal of this section was to compare point estimates for general agreement, given that the GA could not produce error estimates and the formal estimates from LM were too small to be realistic. The NEA lines give both linear and quadratic co-efficients. The NEA radii and inclination values for Kepler-1b -2b, -8b, -12b and -13b are from Esteves~{\em et al.}  (2015), -14b from  Buchinave~{\em et al.} (2011), -15b from Endl~{\em et al.} (2015), and -77b from  Gandolfi~{\em et al.} (2011). We note in passing that Southworth's (2012) values for Kepler 14 are in better agreement with this study than those of Buchinave~{\em et al.} (2011).  These are the only two studies of the planet with inclinations on NEA.

In general the fits by this study were not able to find viable solutions when quadratic limb darkening was introduced as free variables, but could fit solutions when linear limb darkening was included as a free variable.  This is to do with the information content of the data, and is discussed by Ji~{\em et al.} (2017).  For the first two variables both the GA and LM method converged to good agreement with the literature.  However for inclination LM was not in agreement for some of the systems, whereas the GA was in closer agreement.  It appears that the convergence test for the LM method was insufficient for inclination, prematurely ending optimisation, and indicates that particular care needs to be taken when using a single optimisation algorithm in the selection of the stopping conditions. If we had no other method to compare the results against, we might accept that an optimum had been reached.  It also speaks for the use of more sophisticated methods that these which, for instance, explore the $\chi^2$ optimisation space around a declared optimum. Given this, and the general agreement across the models  and optimisation techniques, we proceeded to MCMC models in order to estimate `confidence' in the derived parameters, which would allow more meaningful comparison of the results across  the difference methods and studies. When we used Gaussian Processes to model the out-of-transit light variations, substantially smaller (planetary) radii were estimated for Kepler-14b.  The picture for Kepler-77b was more confused, with WinFitter estimating the planet as smaller but NUTS (in conjunction with a more simple model) more in line with the literature. Given the sophistication of WinFitter, this solution is preferred, and we summarize that the application of Gaussian processes led to smaller estimates for planetary radius.

\begin{table*}[htbp]
\caption{\label{table:mcmc_radial_rhat } {\bf MCMC Radial Velocity  Model Parameter $\hat{R}$ values}}
\centering
\begin{tabular}{llll}
 \hline
Planet & $e$ & $q$ & $RV_C$ \\
 \hline
Kepler-12b & 0.9998  & 1.0002 & 1.0011 \\
Kepler-14b & 1.0000  & 1.0006 & 0.9998 \\
Kepler-15b & 1.0000  & 0.9998 & 1.0000 \\
Kepler-40b & 0.9998 & 0.9997 & 1.0032 \\
51 Peg-b & 1.0029 & 0.9998 & 1.0005 \\
\hline
\label{table:mcmc_radial_rhat}
\end{tabular}
\end{table*}


\begin{figure*}[!htb]
\centering
\subfloat[]{\includegraphics[width=2.5in]{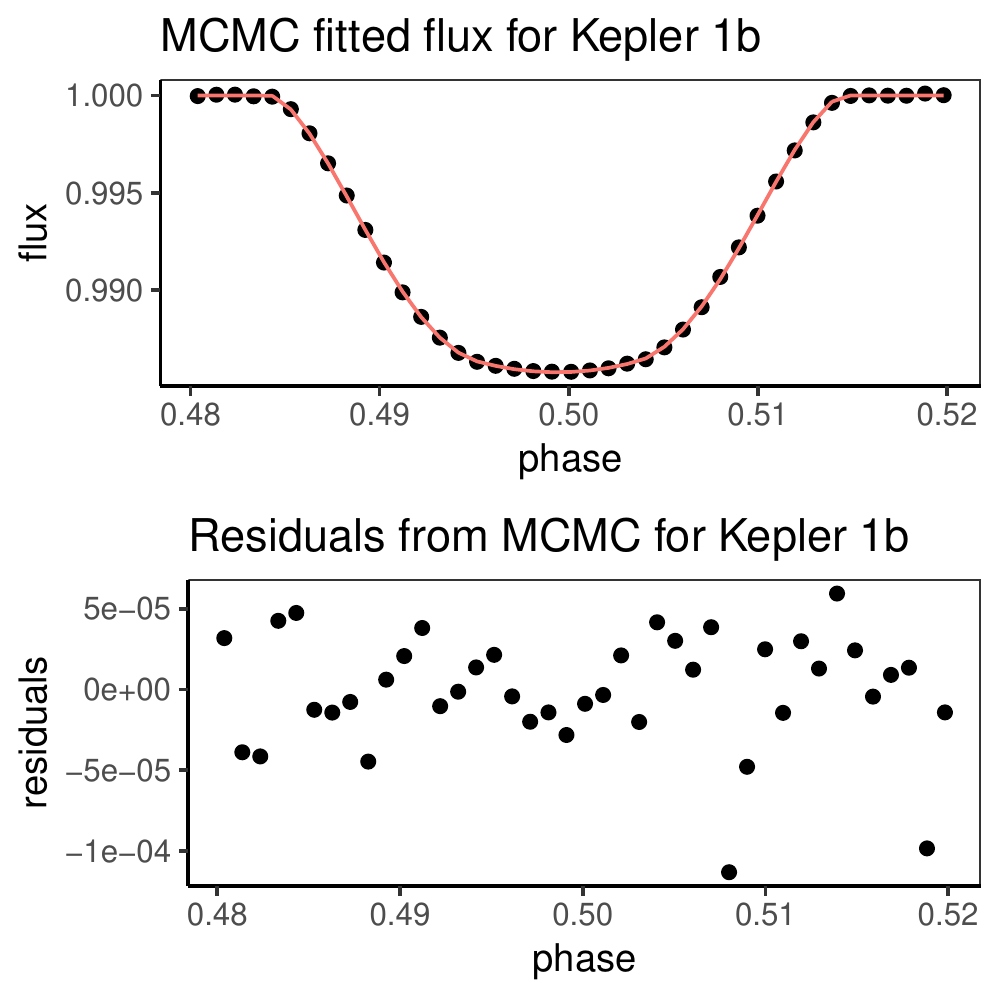}} 
\subfloat[]{\includegraphics[width=2.5in]{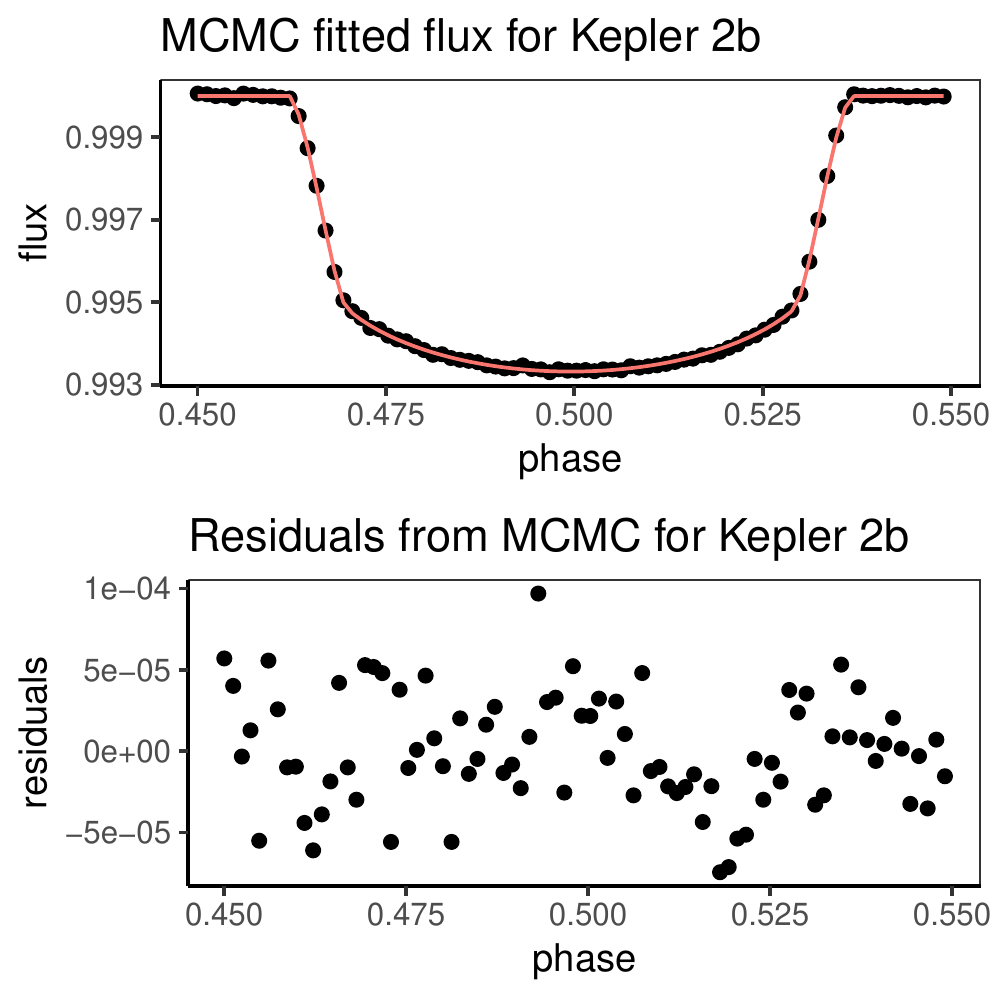}} 
\caption{{\bf Illustrative MCMC Transit Model Fits and Residuals} for Kepler-1b (sub-figure (a) and -2b (b). For the point optimisations we set 50 data points per bin and only looked at data that were near or in the transit. For the later MCMC fits (including EXOFAST) we did not bin the data.
} 
\label{fig:nuts_transit_tests1}
\end{figure*}

\begin{figure*}[!htb]
\centering
{\includegraphics[width=4in]{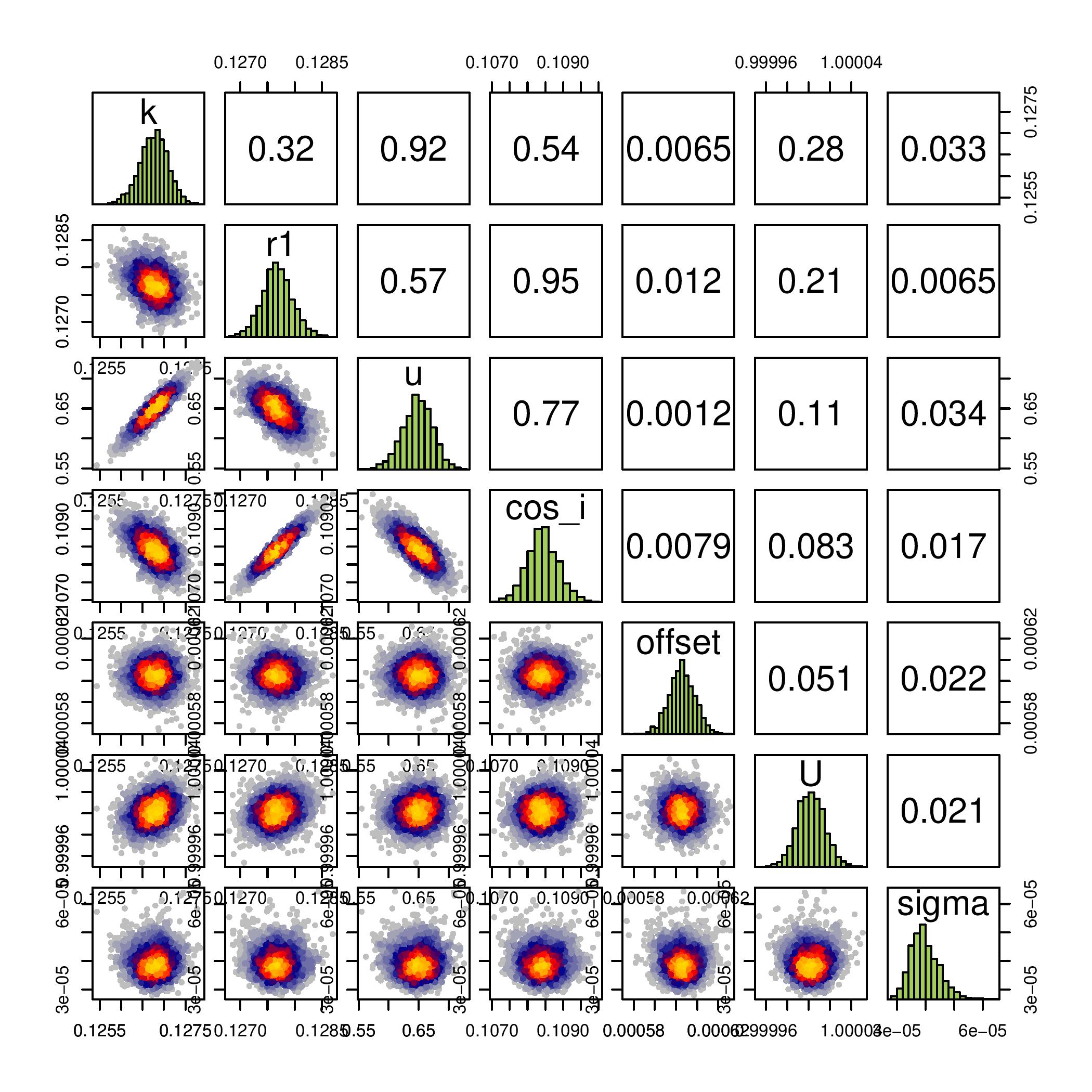}}
\caption{{\bf Example of Transit MCMC Scatter Plot (Kepler-1b).} $k$ is the ratio of the planetary to stellar radii,
`r1' is the planetary radius, $u$ the linear limb darkening coefficient, `cos\_i' the cosine of the inclination, `offset'
the phase offset, $U$ the mean system brightness outside the transit, and `sigma' the random noise. This diagram is representative
of the scatter plots for the other systems modelled in this paper.  Similar charts were produced for the radial velocity fits.
} 
\label{fig:five_kep}
\end{figure*}

\begin{figure*}[!htb]
\centering
\subfloat[]{\includegraphics[width=2in]{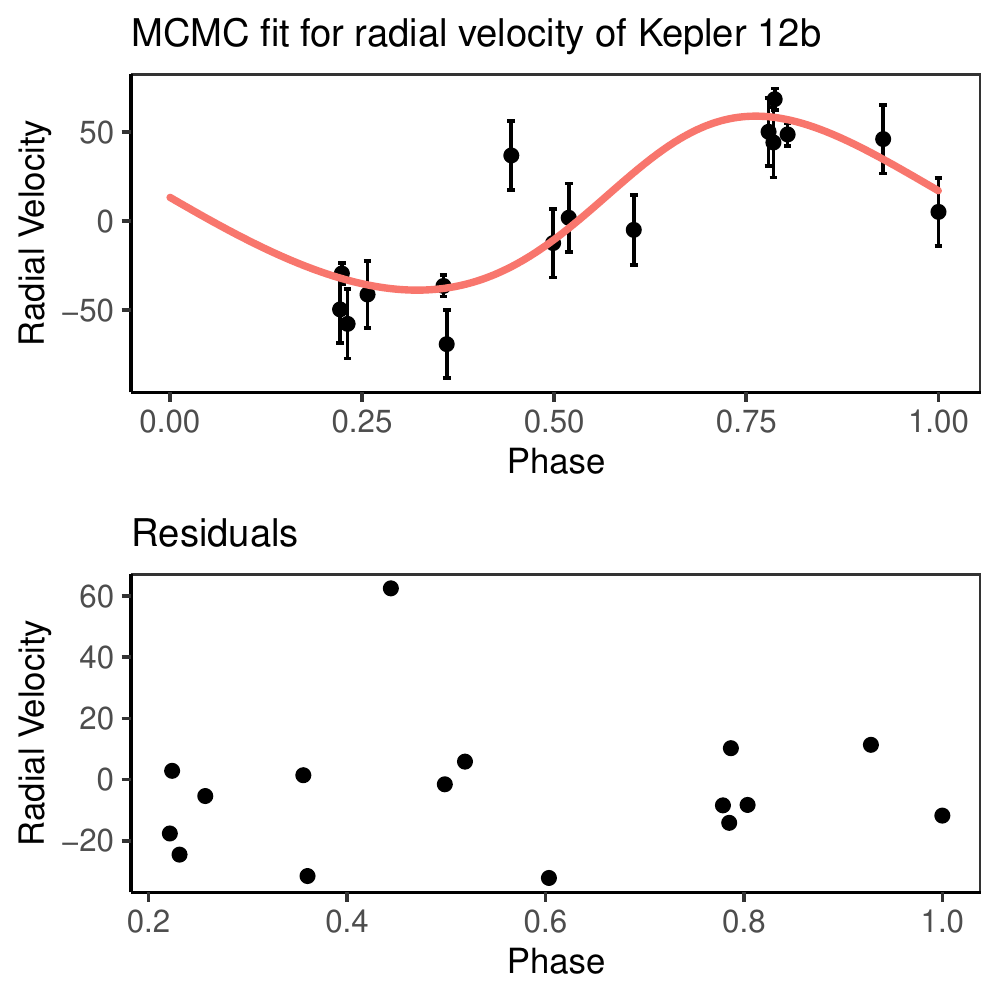}} 
\subfloat[]{\includegraphics[width=2in]{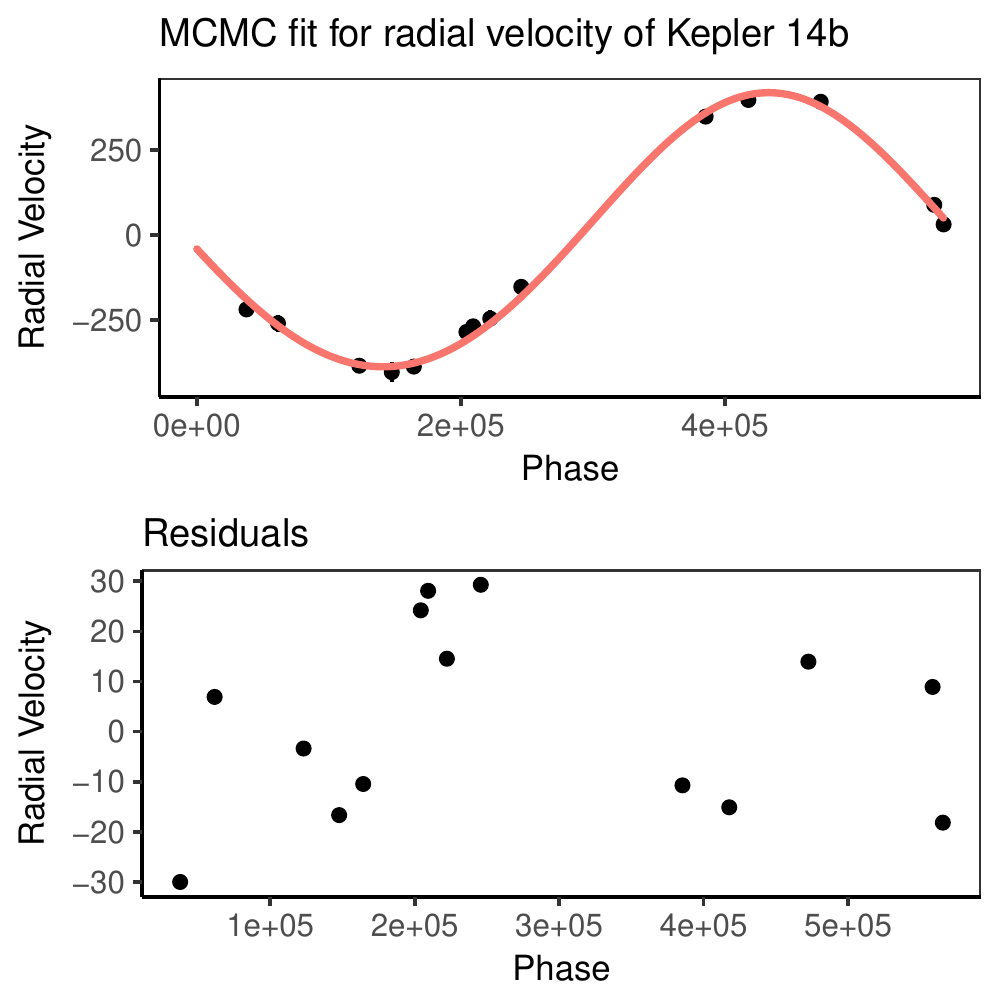}} 
\subfloat[]{\includegraphics[width=2in]{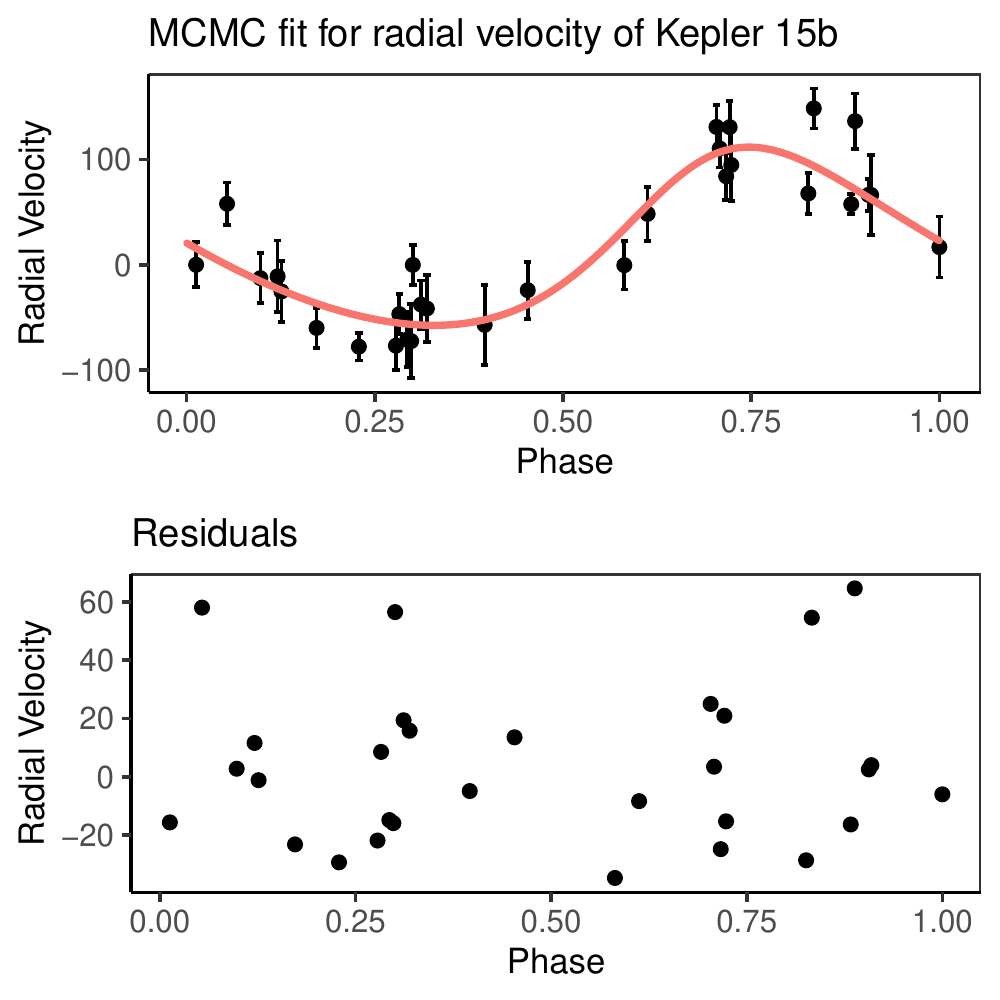}} \\
\subfloat[]{\includegraphics[width=2in]{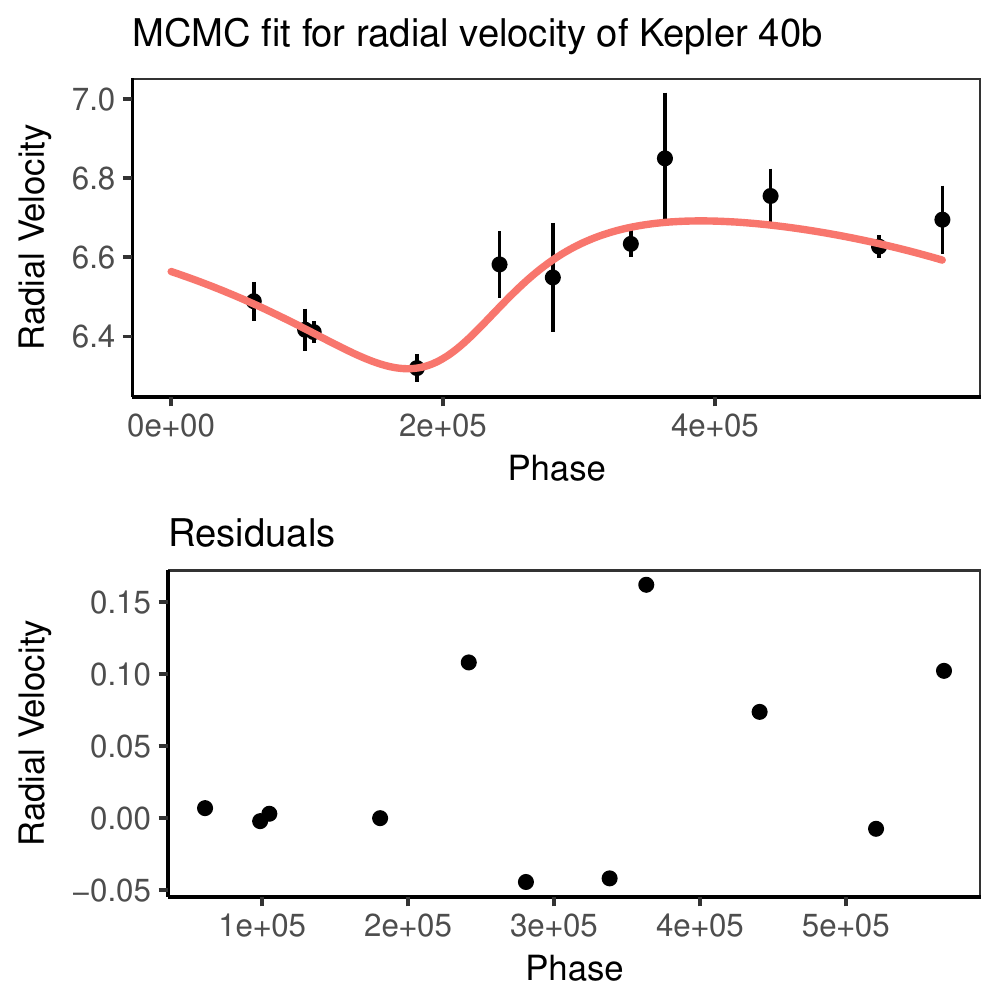}} 
\subfloat[]{\includegraphics[width=2in]{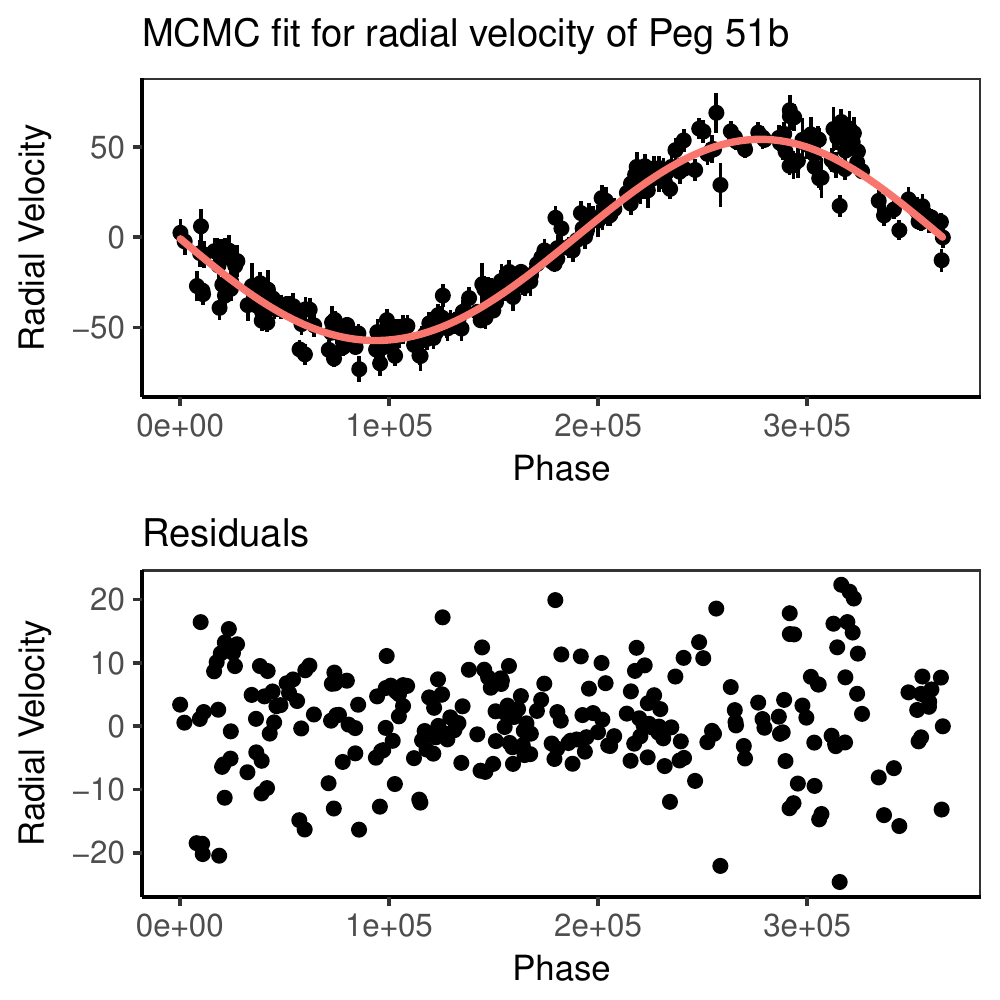}} 
\caption{{\bf STAN MCMC Radial Velocity Model Fits and Residuals.} } 
\label{fig:four_kep}
\end{figure*}


\subsection{The No-U-Turn MCMC Sampler}

To better explore confidence in the parameter estimates, we employed Markov Chain Monte Carlo (MCMC) methods. In particular, we utilised the No-U-Turn Sampler (NUTS)  given that NUTS can automatically tune for parameters that previously required specification by a user. MCMC is more computationally intensive { than} the two optimisation methods used in the previous step. In probability theory, a Markov chain is a sequence of random variables $\theta_1, \theta_2, ...,$ for which, for any $t$, the distribution of $\theta_t$ given all previous $\theta$'s depends only on the most recent value, $\theta_{t-1}$. MCMC is a general method which draws values of $\theta$ from approximate distributions and then improves the draws at each step to better approximate the target posterior distribution, $p(\theta |y)$. The sampling is done sequentially, such that the sampled draws form a Markov chain (see Gelman~{\em et al.}, 2014, as background for this discussion).

In our applications of Markov chain simulation, several independent sequences are created. Each sequence, $\theta_1, \theta_2, \theta_3, ...$, is produced by starting at  some point $\theta_0$. Then, for each $t$, $\theta_t$ is drawn from a transition  distribution, $T_t(\theta_t|\theta_{t-1})$, which depends on the previous draw $\theta_{t-1}$.  Each $\theta_t$ would contain N samples. The transition probability distribution must be  constructed such that the Markov chain converges to a unique stationary distribution  that is the target posterior distribution, $p(\theta|y)$. Convergence was assessed using the $\hat{R}$ statistic defined as (Gelman {\em et al.}, 2014):
\begin{equation}
\hat{R} = \sqrt{\frac{\hat{var}^+ (\psi|y)}{W}},
\end{equation}
which declines to 1 as $n \rightarrow \infty$. We note that:
\begin{equation}
\hat{var}^+ (\psi|y)=\frac{N-1}{N} W +\frac{1}{N} B,
\end{equation}
where $W$ and $B$ refers to the between sample variance estimate and within sample variance estimate respectively. When the chains converge to a stationary distribution, the $\hat{R}$ statistics will converge to 1. This is commonly used as one of the criteria for assessing convergence in MCMC algorithms.

Hoffman \& Gelman (2014) explain the benefits of NUTS succinctly that it  ``...uses a recursive algorithm to build a set of likely candidate points that spans a wide swath of the target distribution, stopping automatically when it starts to double back and retrace its steps. Empirically, NUTS performs at least as efficiently as (and sometimes more efficiently than) a well tuned standard HMC method, without requiring user intervention or costly tuning runs.''  

We made use of the STAN,\footnote{See https://mc-stan.org/ more details. The STAN reference manual is available at https://mc-stan.org/docs/2\_25/reference-manual/index.html} which is a platform accessible to popular data analysis languages such R, Python, MATLAB, Julia, and Stata. STAN provided us with full Bayesian statistical inference with MCMC sampling. It uses an approximate Hamiltonian dynamics simulation based on numerical integration, subsequently corrected by performing a Metropolis acceptance step. The Hamiltonian Monte Carlo algorithm starts at a specified initial set of parameters  
 and then across subsequent iterations a new momentum vector is sampled with the current values of the parameters 
being updated using the leapfrog integrator according to the Hamiltonian dynamics. A  Metropolis acceptance step is applied each iteration, and a decision is made whether to update to the new state or keep the existing state.\footnote{For further details on the algorithm and usage see https://mc-stan.org/docs/2\_25/reference-manual/hamiltonian-monte-carlo.html\#ref-Betancourt-Girolami:2013 and references within for further details.}


\begin{table*}[!htb]
\caption{\bf Density estimates and associated limits for Kepler-12b and -15b from the STAN MCMC fits}
\small
\centering
\begin{tabular}{|c|c|c|c|}
\hline
Exoplanet & Density (\textit{g/cm}$^{3}$) & Lower Limit & Upper Limit \\ \hline
Kepler 12b & 0.096  & 0.059 & 0.155 \\ \hline
Kepler 15b & 0.779 & 0.591 & 1.036 \\ \hline
\end{tabular}
\label{tab:density}
\end{table*}

\subsection{MCMC Results}

Having tested the model produced results in line with the literature, we were ready to move through to MCMC optimisation. Tables~\ref{table:mcmc_transits} , \ref{table:gp_mcmc_transits}, and \ref{table:mcmc_radial} give the NUTS results for the transit and radial velocity fits.  The application of Gaussian processes led to significantly smaller estimates  for  the  planetary  radius  of  Kepler-14b  than  in the literature, while the estimates did not change significantly for Kepler-77b. 

Before optimisation, we performed transformation of some parameters for various reasons. By transforming parameters $R_P$, $R_S$ and $a$ into ratios $k=R_P/R_S$ and $R_S/a$, we were able to reduce the dimensionality of the optimisation problem. These ratios and the limb darkening coefficients $u$, $u_1$ and $u_2$ take values between 0 and 1, which allows utilisation of a uniform prior distribution for sampling draws in MCMC. Similarly, we transformed $i$ into $\cos i$ and used a uniform prior for the parameter. A uniform prior would be ideal for obtaining an unbiased estimate as we assume as little prior information as possible for the optimised parameters. Given the transformed parameters, the optimisation problem to solve would be the following:
\begin{equation}
L_{observed}~\sim \mathcal{N}(L_{fitted}(phase_{observed},\sigma^2)
\end{equation}
where $L_{fitted}$ is the flux given by the final optimised parameters and the observed phase and $\sigma$ is an indicator of noise level in the data. 

Figure~\ref{fig:nuts_transit_tests1} presents the some illustrative transit model fits across the test systems, along with the residuals to those fits, while Table~\ref{table:mcmc_transits} lists the derived parameter values and uncertainties from the fits.  Overall the parameters are in reasonable agreement with those listed from the earlier optimisations given in Table ~\ref{table:single_transit}, although there are some unexpected differences in the inclination estimates for Kepler-12b and -13b. The chains were well converged, as shown by Table~\ref{table:mcmc_transits_rhat}. Figure~\ref{fig:five_kep} shows an example scatter plot of the MCMC results for the optimised parameters.  Tables~\ref{table:mcmc_radial} and \ref{table:mcmc_radial_rhat} list the results for the MCMC fits to radial velocity data, which showed good convergence. By eye, the fits are reasonable (see Figure~\ref{fig:four_kep}) and in reasonable agreement with the previous work given in Table~\ref{table:rv_fits_single}.


\begin{table*}[!htb] 
\caption{\label{table:mcmc_transits_gj357 } {\bf STAN MCMC Transit Model Fit for GJ357-b.}
The mean and one standard deviation errors are given for each optimized parameter, together with the convergence statistic.}
\centering
\begin{tabular}{lll}
\hline
 estimates & mean & $\hat{R}$ \\
 \hline
$k$ & $0.0334 \pm 0.0017$ & 1.0032 \\
$R_S/a$ & $0.0692 \pm 0.0182$ & 1.0020 \\
$u$ & $0.2906 \pm 0.1947$ & 1.0008 \\
$\cos{i}$ & $0.0537 \pm 0.0244$ & 1.0021 \\
\hline
\end{tabular}
\label{table:mcmc_transits_gj357}
\end{table*}


\subsubsection{Kepler-12b}

From the above we were able to calculate the densities for two planets, being present in both the transit and radial velocity fits. These are given in Table~\ref{tab:density}. The mean density for Kepler-12b is similar to those estimates of 0.110 g per cc of Bonomo~{\em et al.} (2017), Esteves~{\em at al} (2015), and Fortney~{\em et al.} (2011) or the 0.108 of Southworth (2012), all of whom gave errors of order 0.01.  To investigate further we ran EXOFAST (Eastman~{\em et al.} (2013), which fits both the transit and radial velocity data together using MCMC. As before, Quarter 2 data were used from Kepler. For Kepler-12b, EXOFAST derived a planetary mass of $0.44 \pm 0.04$ { $M_J$ (Jupiter masses)}, a radius of $1.72 \pm 0.05$ { $R_J$ (Jupiter radii)}, equilibrium temperature of $1490 \pm 30 $  Kelvin, inclination  $88.8^{+0.6}_{-0.4}$ degrees, a Safronov number of $0.024 \pm 0.002$, and a density of $0.11 \pm 0.01$ g per cc.  Errors are all one standard deviations. This density is in better agreement with the literature. EXOFAST derived a much more realistic eccentricity of $0.028^{+0.04}_{-0.02}$ than that above where the radial velocity data were modelled without the transit data. The derived mass ratio was $0.00035 \pm 0.00003$, similar to that from the earlier fit, but the planet radius was slightly smaller at $0.1175^{+0.0005}_{-0.0004}$ relative to its star. This might be due to the model using quadratic limb darkening (coefficients $0.34 \pm 0.02$ and $0.28 \pm 0.04$) compared to the linear term used in the previous model. However, the overall agreement is reassuring.

\subsubsection{Kepler-15b}

Turning to Kepler-15b, we also ran EXOFAST on the same radial velocity data and short cadence Quarter 3 Kepler data finding a solution of $0.73 \pm 0.07$ { $M_J$}, $1.040^{+0.073}_{-0.066}$ { $R_J$}, an equilibrium temperature of $1135^{+44}_{-42}$ Kelvin, a Safronov number of $0.0768^{+0.0098}_{-0.0091}$, an ellipticity of $0.215^{+0.070}_{-0.069}$, inclination of $87.25^{+0.26}_{-0.31}$ degrees, a mass ratio of $0.00066 \pm 0.00006$, and a mean density of $0.81^{+0.19}_{-0.16}$ g per cc. These results are generally close to the STAN MCMC fit results (see Table~\ref{tab:density}), which is again reassuring. Quadratic limb darkening coefficients were $0.45 \pm 0.04$ and $0.21 \pm 0.05$. The mean density in the literature ranges from $ 0.404 \pm 0.048$ (Southworth, 2011) to $0.93 ^{+0.18}_{-0.22}$ (Bonomo~{\em et al}, 2011), placing our density estimates in the upper end of the range.  The ellipticity appears unrealistic, so we reran EXOFAST forcing a circular orbit.  This led to a substantially lower mean estimated density of $0.43 \pm 0.05$ corresponding to final estimates of $0.72 \pm 0.07$ { $M_J$}, $1.27 \pm 0.04$ { $R_J$}, a substantially warmer equilibrium temperature of $1249 \pm 29$ Kelvin, inclination of $86.15^{+0.15}_{-0.14}$ degrees, a smaller Safronov number of $0.058 \pm 0.005$, and a similar mass ratio of $0.00060^{+0.00005}_{-0.00006}$. Limb darkening coefficients were unchanged. The planetary radius relative to its host star was larger than that estimated from the STAN MCMC fit, at $0.0121 \pm 0.0006$.  These results are in good agreement with those of Southworth (2011). Additional radial velocity data would be helpful for this system, better constraining the fits.

\begin{figure*}[!htb]
\centering
\includegraphics[width=3in]{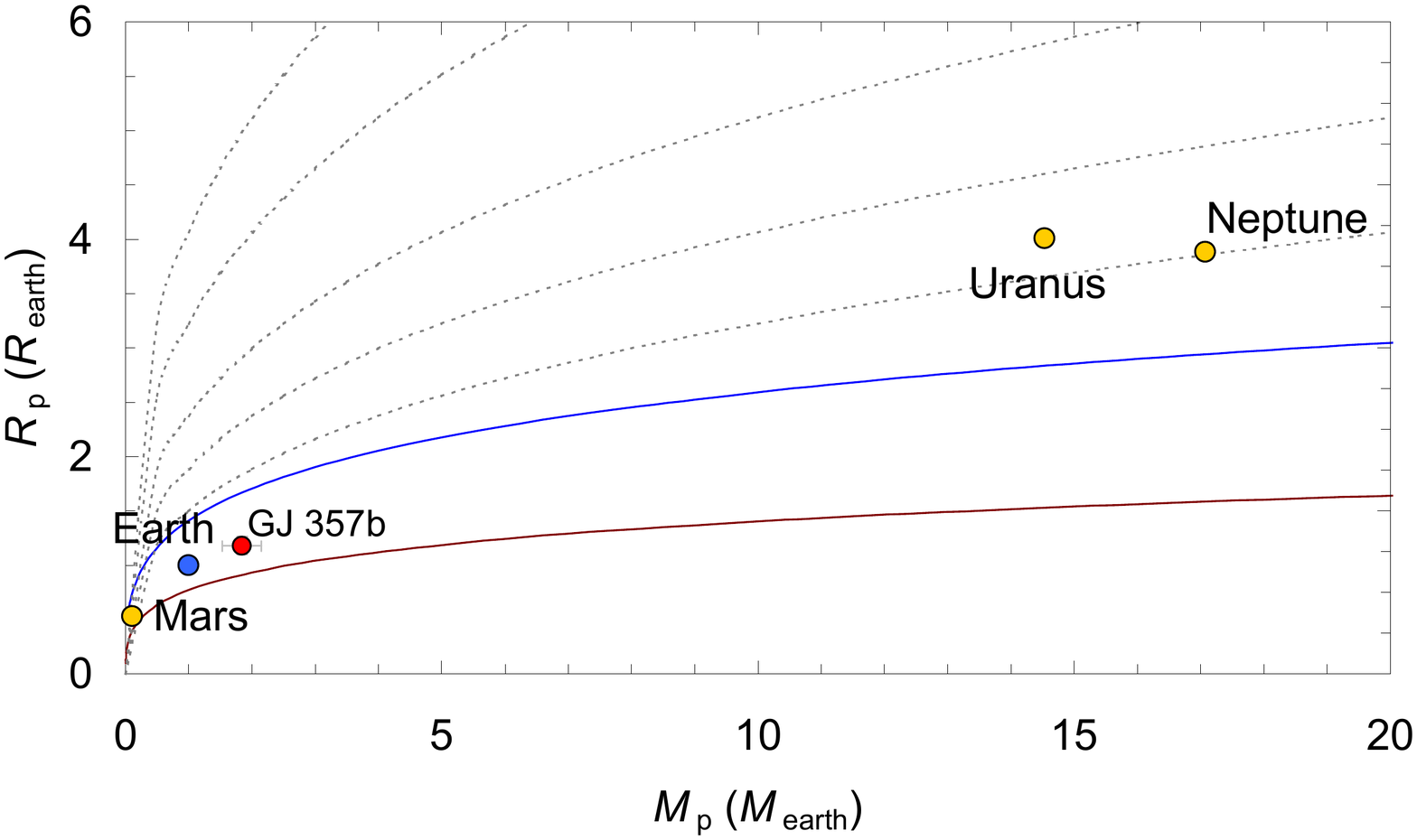} 
\caption{{\bf Density of GJ357b plotted on a mass-radius chart.}
The blue line corresponds to the density of water, and the brown line that of iron. The dashed lines represent 0.05, 0.10, 0.25, 0.50 and 1.00 times the Earth's density respectively.
} 
\label{fig:gj357_density}
\end{figure*}

\begin{table}[!htb]
\begin{center}
\caption{Derived parameters for WinFitter fitting to TESS photometry of GJ 357, as shown in Fig~3.
Parameters for which no error estimate is given are adopted (see Table~\ref{tbl-1b}).
The listed parameters result from combining our HIPPARCOS and ASAS fittings.
\label{tbl-1b}} 
\begin{tabular}{lccc}
   \hline  \\
\multicolumn{1}{c}{Parameter}  & \multicolumn{2}{c}{Value} & 
\multicolumn{1}{c}{Error}\\
\multicolumn{1}{c}{ }  & \multicolumn{1}{c}{WF} & \multicolumn{1}{c}{RWMH} &
\multicolumn{1}{c}{ }\\
\hline \\
$M_p/M_*$ & 0.000016 & --- & --- \\
$L_1$ & 1.0 & --- & ---\\
$L_2$ & 0.0 & ---& --- \\
$R_1 \odot $ & 0.341 & 0.333 & 0.004\\
$R_2 \oplus $ & 1.18  & 1.15 & 0.02 \\
$i$ (deg) & 88.8 & 89.3 & 1.0 \\
$u_1$ & 0.54 & --- & --- \\
$\Delta\phi_0$(deg) &--0.08$^{\circ}$ & 0.0  & 0.2 \\
$\chi^2/\nu$ & 0.96 &  ---  & ---  \\
$\Delta l $& 0.00028&  ---  & --- \\
\hline
\end{tabular}
\end{center}
\end{table}

\begin{table*}[!htb]
\caption{{\bf Prior Input Data  for GJ 357:\label{tbl-1z}}
The notation in the first row is: $L_1$ -- fractional luminosity of host star taking into account companion (see Section 2.1), $M_*$ -- mass of star (solar masses), $R_*$ -- radius of star (solar radii), $T_*$ -- temperature of star (K), $\lambda$ -- effective wavelength ({\AA}), $P$ -- orbital period (in days). The numbers given in the second and fourth rows are rounded averages from results in the cited  literature and sources provided by the NEA (see text). The notation in the third row is: $\bar{\rho_{\star}}$  -- star's mean density (CGS),  $\bar{\rho_p}$ -- planet's mean density, $Z$ -- metallicity of star, $\log g$ -- log$_{10}$ of the surface gravity of star (cgs units), $a$ -- semi-major axis in AUs, $M_p/M_*$ -- ratio of planet to star masses, $u_1$ -- stellar (linear) limb-darkening coefficient, 
}
\label{tbl-1a}
{\footnotesize
\begin{center}
\hspace{2em} 
\begin{tabular}{|r|r|r|r|r|r|r|r|}
\hline 
\multicolumn{1}{|c|}{$L_1$} &
\multicolumn{1}{|c|}{$M_*$} & 
\multicolumn{1}{|c|}{$R_*$} & 
\multicolumn{1}{|c|}{$T_*$} & 
\multicolumn{1}{|c|}{$T_p$} & 
\multicolumn{1}{|c|}{$\lambda$} &
\multicolumn{1}{|c|}{Epoch} & 
\multicolumn{1}{|c|}{$P$ d} \\ 
\hline 
 1.000 & 0.342 &  0.337  & 3500 & 430  & 6220  & 2458517.9994 & 3.93086 \\
 \hline
\multicolumn{1}{|c|}{$\bar{\rho_{\star}}$ } & 
\multicolumn{1}{|c|}{$\bar{\rho_p}$ } & 
\multicolumn{1}{|c|}{$Z$} & 
\multicolumn{1}{|c|}{$\log g$} & 
\multicolumn{1}{|c|}{$a$} & 
\multicolumn{1}{|c|}{ ${M_p}/{M_*}$} & 
\multicolumn{1}{|c|}{ ${R_p}/{R_*}$} & 
\multicolumn{1}{|c|}{$u_1$}  \\
\hline 
 12.6 & 7.8 &  --0.12  & 4.92 & 0.033  & 0.0006  & 0.032 & 0.52 \\ 
 \hline
\end{tabular} 
\end{center}}
\vspace{1ex}
\end{table*} 

\subsubsection{Kepler-14b}

Kepler-14b proved to be a more complex system than we had expected at the beginning of the project.  Bucchave~{\em et al.} (2011) found that the system was in a close visual binary system.  Our fitting (using Kepler Quarter 4 data) did not take into account the dilution effect of the second star, deriving a mean density of $13.9^{+2.7}_{-2.3}$ g per cc.  This is well outside the error limits of Bucchave {\em at al.}, who gave a mean density of $5.7^{+1.5}_{-1.0}$ before the dilution was handled, and $ 7.1 \pm 1.1$ afterwards.  The large eccentricity in our radial velocity fit is not likely realistic, given the close orbit of the planet about its host star. Given that our STAN MCMC fit did not handle the dilution, we have not included our estimate into Table~\ref{tab:density}.  A  first EXOFAST MCMC fit used the undiluted data, deriving a mean density of $5.41^{+0.59}_{-0.51}$ g per cc. This is still far from the STAN MCMC estimate but in good agreement with Bucchave~{\em et al.} In subsequent EXOFAST fits, diluting the flux by the amount reported by Bucchave led to an estimated mean density of $3.12^{+0.35}_{-0.31}$ g per cc, $4.95 \pm 0.17$ { $M_J$}, $1.25 \pm 0.05$ { $R_J$}, an equilibrium temperature of $1596 \pm 27$ Kelvin, eccentricity at $0.023 \pm 0.013$, a Safronov number of $0.42 \pm 0.02$, inclination of $86.2^{+0.05}_{-0.04}$ degrees, a mass ratio of $0.00315 \pm 0.00006$, a planetary radius of $0.0598 \pm 0.0004$ its host star, and a semi-major axis $8.0 \pm 0.2$ the stellar radius.  Limb darkening values were $0.30 \pm 0.03$ and $0.29 \pm 0.03$ respectively. The orbital radius is larger than the STAN estimates, and the density not in agreement with the value of Buchave~{\em et al.}.  We suspected that discrepancies could be due to the effect of `waves' clearly running through the light curve.  A Lomb-Scargle (Lomb, 1976; Scargle, 1982; VanderPlas, 2018) analysis suggested periods of 6.1385, 5.6583, and 4.2557 days. We therefore ran EXOFAST over the Gaussian Process corrected data used earlier, including the dilution.  A circular orbit was assumed given the above results. This led to an estimated density for the planet of $4.04 \pm 0.58$, still lower than those of Bucchave, based on a mass of $4.80^{+0.18}_{-0.17}$ { $M_J$} and $1.137^{+0.069}_{-0.054}$ { $R_J$}. Such a density suggests a rocky composition, perhaps similar to Mars. Modelling the out of transit variations therefore did not resolve the discrepancy.  The other estimated parameters were: effective temperature $1537^{+35}_{-35}$, Safronov number $0.464^{+0.024}_{-0.027}$, orbital semi-major axis (in stellar radii) $8.66^{+0.32}_{-0.34}$, inclination $88.0^{+1.1}_{-0.8}$ degrees, and a mass ratio of $0.00318 \pm 0.00007$. Limb darkening values were $0.271^{+0.037}_{-0.036}$ and $0.294^{+0.046}_{-0.047}$ respectively. We are therefore unable to confirm the densities given by Bucchave~{\em et al.} (2011).

\subsubsection{Kepler-40b}

For completeness, we ran EXOFAST on Kepler-40b Quarter 5 Kepler long cadence data and the radial velocity data used above.  A similar study of {\em Kepler} transit data by Huang~{\em et al} (2019) had discussed the biases introduced in modelling long integration observations and so we were concerned about the effects to our density estimate should we not consider the effect of long integration periods `blurring' out photometric changes.  Kipping (2010) also discussed the problems involved in using long cadence Kepler data, which EXOFAST has attempted to handle, as did Santerne~{\em et al.} when they investigated this planetary system.  We therefore used EXOFAST in preference to our STAN methodology given its handling of long integration periods. 

A circular orbit was fixed (given the sparsity of the radial velocity data), and the long cadence option used in EXOFAST (which resamples the light curve 10 times uniformly spaced over the 29.5 minutes for each {\em Kepler} long cadence data point and averaging).  The optimal solution was for a mass of $2.07 \pm 0.31$ { $M_J$}, a radius of $1.27^{+0.18}_{-0.07}$ { $R_J$}, a mean density of $1.19^{+0.31}_{-0.36}$ g per cc, an equilibrium temperature of $1662^{+85}_{-46}$, a Safronov number of $0.16 \pm 0.03$, and inclination of $87.7^{+1.5}_{-1.9}$ degrees, a planet radius $0.0574^{+0.0011}_{-0.0006}$ that of its host star, and a semi-major axis $7.73^{+0.33}_{-0.74}$ the stellar radius. Quadratic limb darkening values were $0.26 \pm 0.04$ and $0.30 \pm 0.05$ respectively. These are generally in reasonable agreement with the results of Santerne~{\em et al.} (2011), mainly due to the large uncertainties in both study's results. For instance, Santerne~{\em et al.} give the mean planetary density as $1.68^{+0.53}_{-0.43}$ g per cc, also demonstrating large uncertainties for this quantity. 

Kepler-40b is a challenging system to fit, given the accuracy of radial velocity data points (Kepler magnitude $14.58 \pm 0.02$) and long cadence photometry --- total transit duration is estimated at $0.2874^{+0.0035}_{-0.0024}$ days, so one data point is approximate 7\% the duration of the transit. Given that the radial velocity data were obtained with a small telescope (the 1.93-m at Observatoire de Haute Province) it would be interesting for a similar campaign using similar medium aperture telescopes to obtain further such data, which would help firm up the modelling and subsequent results.  Similarly further short integration period (but low noise) photometry would be helpful.



\subsubsection{GJ357}

Photometric data from NASA's Transiting Exoplanet Survey Satellite (TESS) revealed  the Earth-like planet-containing exoplanet system GJ357, as announced in mid-2019 (Luque~{\em et al.}, 2019). GJ357's M-type dwarf star, with mass 0.342 $\pm$ 0.011 M$_{\odot}$, radius 0.337 $\pm$ 0.015 R$_{\odot}$,  and $\sim0.015$ of the solar luminosity, hosts the interesting planet GJ357-b.  This was estimated to be about 20\% larger than the Earth, orbiting { with} a period of about 3.93 d,  at a separation from the star of about 0.033 AU.

With regard to  representative values for the planet's mean surface temperatures $T_p$,  energy balance considerations  lead to  the influx of energy $F_{\rm in}$ absorbed by a planet of radius $R_p$,  given incident mean flux $f_0$,Bond albedo $A_B$ and cross-sectional area $\pi R_p^2$, being :
 \begin{equation}
F_{\rm in} = f_0 (1 - A_B)\pi R_p^2  \,\,\, .
\end{equation}
$A_B$ is zero for a `black body', but from comparison with the familiar cases of Venus and the Earth, 0.72 and 0.39 respectively (Allen, 1974), we set  a prior value of $A_B$ as 0.5.

The radiation emitted from the (spherical) planet $F_{\rm out}$ may then be associated with a representative mean temperature $T_p$, given by (Stefan's Law): 
\begin{equation}
F_{\rm out} = 4\pi R_p^2 \epsilon \sigma T_p^4
\end{equation}
where $\epsilon$ is the emissivity, generally taken to be of order unity, and $\sigma$ is Stefan's constant. In a steady state $F_{\rm in} = F_{\rm out}$, and so
\begin{equation}
T_p = (f_0[1 - A_B]/4 )^{1/4}
\end{equation}
Continuity of the stellar flux $f_0$ from the source then allows
\begin{equation}
T_p \approx T_{\star} (1 - A_B)^{1/4} \sqrt{r_1/2}  \,\,\,  ,
\end{equation}
where $T_{\star}$ is the star's effective surface temperature. The average surface temperature estimated in this way for GJ357-b  is about 430 K.  While GJ357-b thus lies essentially outside the `habitable zone' (HZ),  the object gains attention as the third-nearest transiting exoplanet yet known, and potentially   suitably arranged for the study of rocky planet composition.  

In the course of examination of supporting spectroscopic observations, Luque {\em et al.} (2019) found two additional planetary candidates in the system. The more separated of these (GJ357-d) orbits within the HZ with a $\sim$56 d period. Depending on its mass, which is still not well-established, but estimated at around 6 M$_{\rm Earth}$, GJ357-d could retain a sufficient amount of atmosphere to support life-like biochemical  processes.  The other planet, GJ357-c, has a mass of at least 4 M$_{\rm Earth}$, orbital period $\sim$9.125 d, separation $\sim$0.061 AU,  and mean temperature that has been estimated to be about 400K, implying a low Bond albedo.   The object has not been confirmed photometrically (nor has GJ357-d, which could be due to their orbital inclinations not leading to transits), though it should have been detected if its orbit were within $\sim 1.5^{\circ}$ of 90$^{\circ}$, which compares with the $\sim 88.5^{\circ}$ found for GJ357-b  

We also sourced TESS photometry for GJ357, along with radial velocity data from the literature following the interesting work of Luque~{\em et al.} (2019). While only one of these planet candidates transits, it was classified as a `super-earth', making it substantially smaller than the other planets studied in this paper and an interesting analysis challenge. 

Firstly we modelled these data using the STAN MCMC methodology. Table~\ref{table:mcmc_transits_gj357} lists the MCMC results for GJ357-b. While the chains are clearly converged, the estimated errors are substantial (approximately 5\% of $k$, 26\% for the orbital radius, and nearly 3 degrees from maximum to minimum inclination). Limb darkening was not surprisingly poorly estimated. {\sc WinFitter} estimates are given in Table~\ref{tbl-1b} based on the same TESS data indicate a planet radius of $ R_2 = 1.18 \pm 0.02 ~R_{\oplus} $,  stellar radius of $ R_1 = 0.341 \pm 0.004 ~R_{\odot} $, and an inclination of $ i = 88.8 \pm 1.0$ degrees, with an eccentricity of $ e = 0.278 \pm 0.053 $.  {\sc WinFitter} prior information is given in Table~\ref{tbl-1z}. The comparison between the parameters estimated from the STAN MCMC fit (Table~\ref{table:mcmc_transits_gj357}) with those from {\sc WinFitter} are good, and also in line with Luque~{\em et al.} (2019). 

{\sc WinFitter} can model radial velocity data, as well as transit data.   When we used all of the radial velocity data modelled by Luque~{\em et al}, we were unable to calculate a density for the transiting planet. We were unable to find a solution to the radial velocity data, { nor} could we confirm the periods/existence of the two non-transiting planets proposed by Luque~{\em et al.} This was despite realigning the subsets via their median or mean radial velocities and following an iterative pre-whitening and period analysis similar to that performed by Luque~{\em et al.} Neither {\sc WinFitter} nor the STAN MCMC methods could find determinate solutions. We therefore used the derived ({\sc WinFitter}) planet radius and the mass from Luque~{\em et al.} to calculate a bulk density of $ \rho = 6.15~gcm^{-3} $.   This value would confirm GJ357-b as being a Earth-like rocky planet, locating it between water and iron density lines in a mass-radius diagram (see Figure~\ref{fig:gj357_density}).

\begin{figure*}[!htb]
\centering
\includegraphics[width=3.5in]{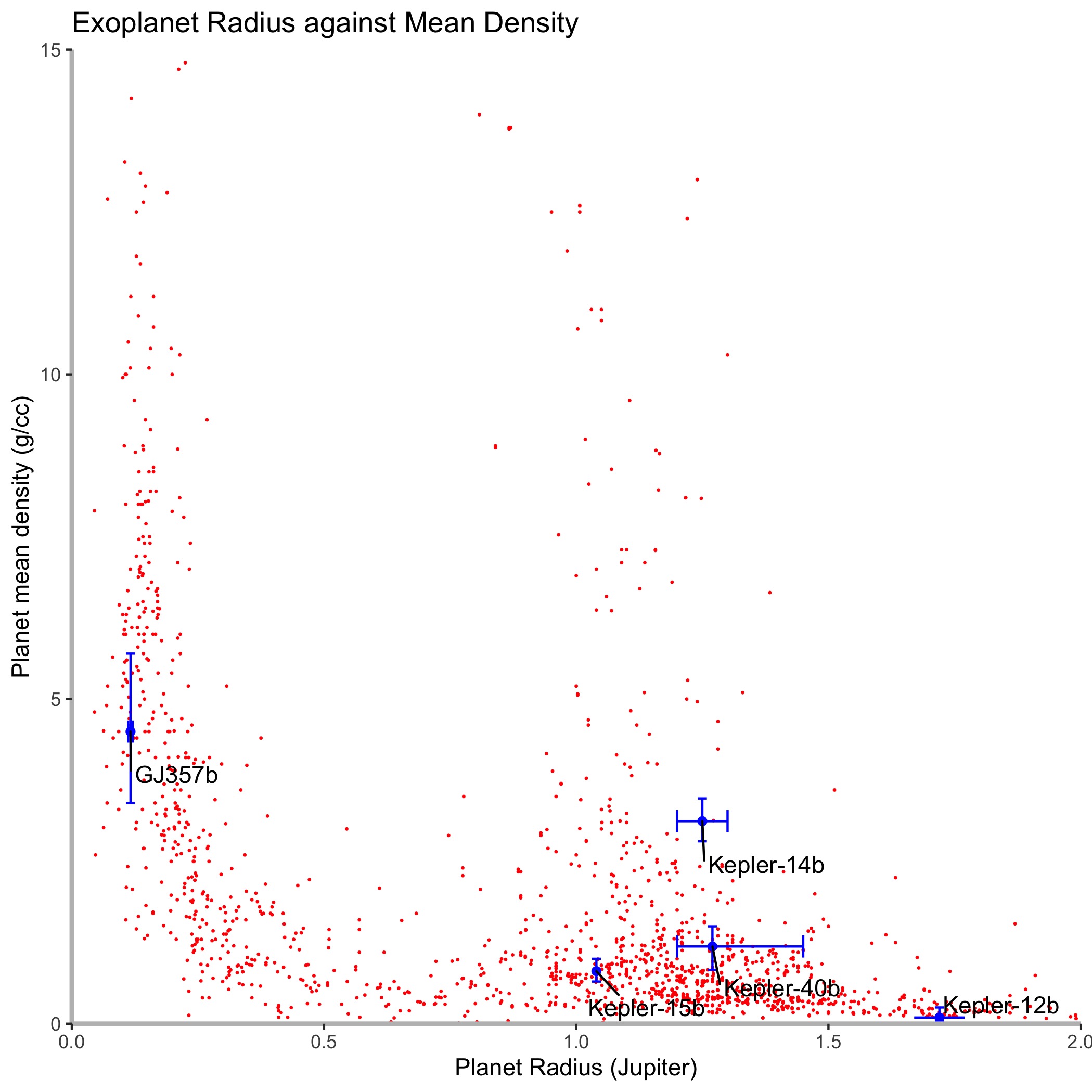} 
\caption{{\bf Exoplanet Radii and Mean Densities}
taken from the NEA database (as of { July 2021}), with the estimates from this paper included.  Error bars were not included from the NEA database as these are incomplete and varying in quality.  There are some interesting features in this chart, such as a possible peak in densities around one Jupiter radius and the apparent increase in densities at low radii.  The second will partly be a selection effect (combined with a likely increase in density towards the host star) given that for a given volume, high densities will produce a more easily observable radial velocity. 
} 
\label{fig:NEA_all}
\end{figure*}

 However we were able to find a radial velocity solution when we used a subset of those data, namely the HIRES and UVES data sets and using EXOFAST.  {\sc WinFitter} was not able to find solutions when the mass ratio $q$ was included as a free parameter. The results of the combined EXOFAST transit and radial velocity analysis are given as { Table~\ref{tab:exofast.}.}  A circular orbit was assumed. The derived density $4.5_{-1.1}^{+1.3}$ g per cc is somewhat lower than the $5.6^{+1.7}_{-1.3}$ g per cc of Luque~{\em et al.}, but well within the (combined) derived uncertainties. The planet radius and mass are in good agreement with Luque.  The insolation is some 17.6 times that on the Earth's surface from the Sun, contributing towards the high equilibrium temperature near six hundred Kelvin, which would place the planet's likely surface temperature well outside the range of habitability as we understand it.  We caution, however, that the fit requires extremely accurate radial velocity measurements and that the results should be treated with caution. The mean error across the radial velocity measurements is 3.2 m/s, of similar magnitude to the radial velocity `signal'.  
 
 A suitably selected subsection of the data gives a plausible model (especially using the photometry), but we cannot unshakably confirm the complete set of previously published results. We next employed Data \& Analysis Center for Exoplanets (DACE) to analyse the same radial velocity data sets analysed by Luque~{\em et al.} (bar PFSpre which was not included as these data `diluted' the periodogram based on all the data), from their source publications. DACE is a facility based at the University of Geneva (CH) dedicated to extrasolar planets data visualisation, exchange and analysis. The Keplerian model initial conditions in DACE are computed using the formalism described in Delisle {\em et al.} (2016),  the analytical FAP values are estimated following Baluev (2008), the numerical FAP values are computed by permutation of the calendar, and the MCMC algorithm is described in Diaz {\em et al.} (2014, 2016). Three signals were detected and modelled in turn as Keplerian orbits, starting at 54.83 days, then 9.12 days, and finally 3.93 days.  These correspond well with the periods found by Luque~{\em et al.}, which were 3.93, 9.12, and 55.66 days. A 500,000 step MCMC chain led to one standard deviation estimates for the period of GJ357-b as [3.929-3.930] days, [9.126-9.127] for GJ357-c, and effectively a point estimate of 54.830 days for GJ357-d.  The orbital distances for each planet in turn were [0.047915-0.049542], [0.08402-0.08687],  and [0.27766-0.28709] AU.  These do not overlap with the estimates by Luque, being larger. $\omega$ and $\lambda$ were poorly constrained, typically ranging over some 180 degrees.  Estimated eccentricities were low (but overall not well constrained) at [0.039-0.290], [0.043-0.262], and 0.250 respectively.  We reran the MCMC analysis forcing circular orbits, given the above results, finding periods of [3.930018-3.930029], [9.12480-9.12486], and [54.8712-54.8728] days for the three suspected planets, leading to mass estimates ($m \sin{i}$) of [1.43-1.74], [2.02-2.41], and [4.89-5.66] Earths respectively. By comparison, Luque~{\em et al.} give mass estimates of $1.84^{+0.31}_{-0.31}$, $> 3.40^{+0.46}_{-0.46}$, and $> 6.1^{+1.0}_{-1.0}$ Earth masses. The orbital period for GJ357-d led to a possible transit for this `planet' being outside the TESS data collection period, while a possible transit for GJ357-c was inside this period.  We could not find this transit in the photometric data.   We did not have stellar activity measures from the literature sources to be able to distinguish such activity cycles from suspected orbital periods. While we can confirm that there are three periods in the literature radial velocity data sets, when they are `cleaned' in the sequence given above, further high accuracy data are needed to confirm the two candidate planets (c and d). The DACE results are tantalising but not conclusive, further observations are clearly needed.


\section{Discussion}

We have built from first principles simple models for transit and radial velocity curve fitting, applying these to a number of systems first using simple optimisation techniques before moving to MCMC. Our findings are in broad agreement with literature results. In the cases of systems with both radial velocity and transit data we were able to make estimates of mean density (see Figure~\ref{fig:NEA_all} for a comparison of the derived densities against densities from the NEA database).  { It is reassuring that the derived densities are consistent with the wider literature.}

The faintness of the {\em Kepler} systems, combined with the intrinsically small radial velocities induced by the planets, makes the collection of well-distributed radial velocity data difficult.  We have shown that deriving estimates of mean density is challenging given other problems, including blended light (Kepler-40b), long integration times for photometry (also Kepler-40b), the low radial velocity amplitude that is comparable to the statistical measurement noise (GJ357-b),  and photometric variability of the host star (Kepler-15b). Sparsity of radial velocity data led to what are likely spurious estimates of orbital eccentricity (such as for Kepler-12b, -15b, and -40b). 

Our derived densities from multiple methods are in reasonable internal agreement (STAN MCMC, {\sc WinFitter}, and EXOFAST).  We caution against taking the parametrization of GJ357, in particular, at face value, in view of the significant scale of data uncertainty and urge additional observations of this system for better future modelling. We trust that the substantial analysis and comparisons made of other systems in this paper, ahead of the GJ357 analysis, support that validity of the methods applied to the system and our conclusions for it. 

We note the comments by Dorn~{\em et al.} (2018) on the current scarcity of low mass exoplanet density estimates and their large uncertainties. ``Currently there are a few dozen super-Earths with measured mass and radius, but only ten or so have mass and radius uncertainties below 20\%''. Characterizing exoplanet masses clearly becomes more challenging as research moves towards smaller masses.   We have shown in this paper that even in `uncomplicated' systems, such as those covered here, the more basic parameter estimates, such as planetary radii, can vary substantially between studies --- often beyond the formally quoted uncertainties, { with published uncertainties for many studies being over-optimistic (see Fig~\ref{fig:nea_comparison} for a sample comparison).}  This raises concerns about propagated errors, { a factor we do not consider our study to be immune to.}
With this, we would stress that realistic total error budgets, { or careful assessment of errors (see below)},  are required for meaningful assessment of the large-sample studies that exoplanet science is moving into.  

{ The Total Error (TE, sometimes also called Total Analytical Error or TAE), represents the overall error in an analysis or statistical test.  This is attributed to systematic and random causes across the entire data collection and analysis process (see, e.g., Oosterhuis \& Theodorsson, 2016). These may be `natural' in cause, such as the intrinsic variability of exoplanet host stars such as plages and rotating spots (Rajpaul et al, 2021). Other effects might include instrument stability or drift, as well as factors in the analysis software.  For instance, different limb-darkening `rules' might be employed; some models employ numerical methods to model distortions (e.g., Wilson \& Devinney, 1971) while others use algebric models (such as WinFitter). 

Many data analysis techniques applied to exoplanets are based on the same formulae as those described in this paper (such as the equations of Mandel \& Agol, 2002) and similar optimisation algorithms (such as variants of MCMC) that contribute towards general agreement between studies.  However between such studies there are clearly also other factors leading to the lack of full agreement (as shown Fig~\ref{fig:nea_comparison}). We would recommend carefully curating data and judicious application of standard processes.  We note that the more advanced fitting functions used in this study are in 
a broad agreement with the other methods.  However, the high quality of current and future data
require that we move from general to a formal closeness of results with
appropriate estimates of fundamental parameters and their uncertainties.

Previously the field of binary stars grappled with similar issues to those faced by the exoplanet community today, namely comparability of results using different analysis techniques (see, e.g., Banks \& Budding, 1991) and treatment of total errors.  No single light curve model or analysis technique was adopted by researchers, given the overall philosophy that science advances through independent confirmations. But there is agreement about careful curation of data. This point is made in Torres et al.'s (2010) painstaking examination of stellar masses and radii based on 95 systems they assessed as being of sufficient precision. Torres et al.\ recomput\-ed the masses and radii, using up-to-date physical constants, to ensure uniformity of findings.  This overall approach is desirable for
exoplanets parameters, given the increasing number of studies and available results. 
Transparency on techniques and methods in the source information is essential for such meta-analyses.
We would do well to emulate Daniel Popper's careful work on binary stars, referred to in his obituary with ``contrary to much astronomical custom, Popper attached, generously realistic error bars to his measurements, so that more recent redeterminations have confirmed his early results more precisely than could reasonably have been expected'' (Trimble, 1999).
}

We eagerly await the outcome of projects like EXPRESS (Jurgenson {\em et al.}, 2016) and EXPRESSO (Hernandez {\em et al.}, 2018) and the public release of their data. This will allow independent follow-up studies like this one, and hopefully reduce the ambiguities shown here in exoplanet modelling via higher cadence and precision radial velocities, { leading to definitive, curated studies as discussed above.}


\section{Acknowledgements}

It is a pleasure to thank Prof.\ Osman Demircan and the colleagues in the Physics Department of COMU (\c{C}anakkale, Turkey) for their interest and support of this programme. The research has been supported by T\"{U}B\.{I}TAK (Scientific and Technological Research Council of Turkey) under Grant No.\ 113F353.  This paper includes data collected by the Kepler mission and obtained from the MAST data archive at the Space Telescope Science Institute (STScI). Funding for the {\em Kepler} mission is provided by the NASA Science Mission Directorate. STScI is operated by the Association of Universities for Research in Astronomy, Inc., under NASA contract NAS 5–26555. This research has made use of the NASA Exoplanet Archive, which is operated by the California Institute of Technology, under contract with the National Aeronautics and Space Administration under the Exoplanet Exploration Program;  EXOFAST (Eastman~{\em et al.} 2013) as provided by the NASA Exoplanet Archive; and {\em Lightkurve}, a Python package for Kepler and TESS data analysis.  The publication makes use of the Data \& Analysis Center for Exoplanets (DACE), which is a facility based at the University of Geneva (CH) dedicated to extrasolar planets data visualisation, exchange and analysis. DACE is a platform of the Swiss National Centre of Competence in Research (NCCR) PlanetS. The DACE platform is available at https://dace.unige.ch. Additional help and encouragement for this work has come from the National University of Singapore (see Ng, 2019), particularly through Prof. Lim Tiong Wee of the Department of Statistics and Applied Probability.  { We thank the University of Queensland for support via collaboration software, and the anonymous referee for their careful and helpful comments which improved this paper.}

\pagebreak
{\onecolumn
\begin{deluxetable}{lcl}
\tablecaption{EXOFAST median values and 68\% confidence intervals for GJ357b.} \label{tbl:exofast_gj357b}
\tablehead{\colhead{~~~Parameter} & \colhead{Units} & \colhead{Value}}
\startdata
\sidehead{Stellar Parameters:}
                           ~~~$M_{*}$\dotfill &Mass (\msun)\dotfill & $0.514_{-0.042}^{+0.043}$\\
                         ~~~$R_{*}$\dotfill &Radius (\rsun)\dotfill & $0.377_{-0.012}^{+0.011}$\\
                     ~~~$L_{*}$\dotfill &Luminosity (\lsun)\dotfill & $0.0275_{-0.0044}^{+0.0046}$\\
                         ~~~$\rho_*$\dotfill &Density (cgs)\dotfill & $13.50_{-0.60}^{+0.63}$\\
              ~~~$\log(g_*)$\dotfill &Surface gravity (cgs)\dotfill & $4.995\pm0.018$\\
              ~~~$\teff$\dotfill &Effective temperature (K)\dotfill & $3830_{-120}^{+110}$\\
                              ~~~$\feh$\dotfill &Metalicity\dotfill & $0.02_{-0.16}^{+0.14}$\\
\sidehead{Planetary Parameters:}
                              ~~~$P$\dotfill &Period (days)\dotfill & $3.93048_{-0.00015}^{+0.00014}$\\
                       ~~~$a$\dotfill &Semi-major axis (AU)\dotfill & $0.0390\pm0.0011$\\
                             ~~~$M_{P}$\dotfill &Mass (\mj)\dotfill & $0.0057_{-0.0013}^{+0.0016}$\\
                           ~~~$R_{P}$\dotfill &Radius (\rj)\dotfill & $0.1164_{-0.0045}^{+0.0044}$\\
                       ~~~$\rho_{P}$\dotfill &Density (cgs)\dotfill & $4.5_{-1.1}^{+1.3}$\\
                  ~~~$\log(g_{P})$\dotfill &Surface gravity\dotfill & $3.02\pm0.11$\\
           ~~~$T_{eq}$\dotfill &Equilibrium Temperature (K)\dotfill & $573_{-18}^{+17}$\\
                       ~~~$\Theta$\dotfill &Safronov Number\dotfill & $0.0075_{-0.0017}^{+0.0021}$\\
               ~~~$\fave$\dotfill &Incident flux (\fluxcgs)\dotfill & $0.0246\pm0.0030$\\
\sidehead{RV Parameters:}
                    ~~~$K$\dotfill &RV semi-amplitude (m/s)\dotfill & $1.15_{-0.26}^{+0.32}$\\
                 ~~~$M_P\sin i$\dotfill &Minimum mass (\mj)\dotfill & $0.0057_{-0.0013}^{+0.0016}$\\
                       ~~~$M_{P}/M_{*}$\dotfill &Mass ratio\dotfill & $0.0000107_{-0.0000024}^{+0.0000030}$\\
               ~~~$\gamma$\dotfill &Systemic velocity (m/s)\dotfill & $0.76_{-0.41}^{+0.40}$\\
\sidehead{Primary Transit Parameters:}
                ~~~$T_C$\dotfill &Time of transit (\bjdtdb)\dotfill & $2458518.00039_{-0.00059}^{+0.00071}$\\
~~~$R_{P}/R_{*}$\dotfill &Radius of planet in stellar radii\dotfill & $0.03174_{-0.00074}^{+0.00072}$\\
     ~~~$a/R_{*}$\dotfill &Semi-major axis in stellar radii\dotfill & $22.25\pm0.34$\\
              ~~~$u_1$\dotfill &linear limb-darkening coeff\dotfill & $0.372\pm0.059$\\
           ~~~$u_2$\dotfill &quadratic limb-darkening coeff\dotfill & $0.321_{-0.057}^{+0.056}$\\
                      ~~~$i$\dotfill &Inclination (degrees)\dotfill & $89.13_{-0.13}^{+0.14}$\\
                           ~~~$b$\dotfill &Impact Parameter\dotfill & $0.336_{-0.050}^{+0.047}$\\
                         ~~~$\delta$\dotfill &Transit depth\dotfill & $0.001007\pm0.000046$\\
                ~~~$T_{FWHM}$\dotfill &FWHM duration (days)\dotfill & $0.05295_{-0.00078}^{+0.00075}$\\
          ~~~$\tau$\dotfill &Ingress/egress duration (days)\dotfill & $0.001896_{-0.000076}^{+0.000083}$\\
                 ~~~$T_{14}$\dotfill &Total duration (days)\dotfill & $0.05485_{-0.00077}^{+0.00073}$\\
      ~~~$P_{T}$\dotfill &A priori non-grazing transit prob\dotfill & $0.04351_{-0.00065}^{+0.00067}$\\
                ~~~$P_{T,G}$\dotfill &A priori transit prob\dotfill & $0.04637_{-0.00070}^{+0.00072}$\\
                            ~~~$F_0$\dotfill &Baseline flux\dotfill & $1.000129\pm0.000017$\\
\sidehead{Secondary Eclipse Parameters:}
              ~~~$T_{S}$\dotfill &Time of eclipse (\bjdtdb)\dotfill & $2458519.96563_{-0.00054}^{+0.00065}$
\enddata
\label{tab:exofast.}
\end{deluxetable}
}

\twocolumn


 \end{document}